\pdfoutput=1

\RequirePackage{fix-cm}

\documentclass[twocolumn]{svjour3}

\smartqed

\usepackage{graphicx}
\usepackage{amssymb,amsmath}
\usepackage{txfonts}
\usepackage{natbib}
\usepackage{hyperref}

\journalname{Astrophysics and Space Science}

\begin{document}

\title{Basins of convergence of equilibrium points \\
in the pseudo-Newtonian planar circular \\
restricted three-body problem}

\author{Euaggelos E. Zotos}

\institute{Department of Physics, School of Science, \\
Aristotle University of Thessaloniki, \\
GR-541 24, Thessaloniki, Greece\\
Corresponding author's email: {evzotos@physics.auth.gr}}

\date{Received: 5 August 2017 / Accepted: 12 September 2017}

\titlerunning{Basins of convergence in the pseudo-Newtonian planar circular restricted three-body problem}

\authorrunning{Euaggelos E. Zotos}

\maketitle

\begin{abstract}

The Newton-Raphson basins of attraction, associated with the libration points (attractors), are revealed in the pseudo-Newtonian planar circular restricted three-body problem, where the primaries have equal masses. The parametric variation of the position as well as of the stability of the equilibrium points is determined, when the value of the transition parameter $\epsilon$ varies in the interval $[0,1]$. The multivariate Newton-Raphson iterative scheme is used to determine the attracting domains on several types of two-dimensional planes. A systematic and thorough numerical investigation is performed in order to demonstrate the influence of the transition parameter on the geometry of the basins of convergence. The correlations between the basins of attraction and the corresponding required number of iterations are also illustrated and discussed. Our numerical analysis strongly indicates that the evolution of the attracting regions in this dynamical system is an extremely complicated yet very interesting issue.


\end{abstract}

\section{Introduction}
\label{intro}

Undoubtedly, one of the most intriguing as well as important fields in dynamical astronomy and celestial mechanics is the few-body problem and especially the version of the circular restricted three-body problem \citep{S67}. This is true if we take into account that this problem has numerous applications in many research fields, such as molecular physics, chaos theory, planetary physics, or even stellar and galactic dynamics. This is exactly why this topic remains active and stimulating even today.

For describing, in a more realistic way, the motion of massless particles in the Solar System several modifications of the classical three-body problem have been proposed, mainly by adding perturbing terms to the effective potential. The classical Newtonian three-body problem is just a first good approximation of a much more complex setting. On this basis, additional general relativistic corrections must be included in order to refine our current understanding of the Solar System dynamics.

In this vein, the first-order post-Newtonian equations of motion for the circular restricted three-body problem have been derived \citep[e.g.,][]{B72,C76,MD94,K67}, by using the Einstein-Infeld-Hoffmann theory \citep{EIH38}. Recent studies indicate that the additional post-Newtonian terms act as non-negligible perturbations to the classical system \citep{DLCG17b}. Especially, when the distance between the two primary bodies is sufficiently small the post-Newtonian dynamics substantially differ from the corresponding classical Newtonian dynamics \citep{HW14}.

Knowing the basins of convergence, associated with the libration points, is an issue of great importance, since the attracting domains reflect some of the most intrinsic properties of the dynamical system. For obtaining the basins of attraction one should use an iterative scheme (i.e., the Newton-Raphson method) and scan a set of initial conditions in order to reveal their final states (attractors). Over the past years a large number of studies have been devoted on determining the Newton-Raphson basins of convergence in many types of dynamical systems, such as the Hill's problem \citep[e.g.,][]{D10}, the Sitnikov problem \citep[e.g.,][]{DKMP12}, the restricted three-body problem with oblateness and radiation pressure \citep[e.g.,][]{Z16}, the photogravitational Copenhagen problem \citep[e.g.,][]{K08}, the electromagnetic Copenhagen problem \citep[e.g.,][]{KGK12,Z17b}, the four-body problem \citep[e.g.,][]{BP11,KK14,Z17a}, the photogravitational four-body problem \citep[e.g.,][]{APHS16}, the ring problem of $N + 1$ bodies \citep[e.g.,][]{CK07,GKK09}, or even the restricted 2+2 body problem \citep[e.g.,][]{CK13}.

In this work we shall use the numerical methodology introduced in \citet{Z16} in order to investigate the dynamics of the pseudo-Newtonian planar circular restricted three-body of \citep{DLCG17a}. The present article has the following structure: the most important properties of the mathematical model are presented in Section \ref{mod}. The parametric evolution of the position as well as of the stability of the equilibrium points is investigated in Section \ref{param}. The following Section contains the main numerical results, regarding the evolution of the Newton-Raphson basins of convergence. Our paper ends with Section \ref{conc}, where we emphasize the main conclusions of this work.

\section{Properties of the mathematical model}
\label{mod}

Let us briefly recall the most important aspects of the circular restricted three-body problem. The two primary bodies, $P_1$ and $P_2$ move on circular orbits around their common center of mass, according to the theory of the classical restricted three-body problem \citep{S67}. It is assumed that the mass of the third body $m$ is significantly smaller with respect to the masses of the primaries ($m \ll m_1$ and $m \ll m_2$). Therefore the third body acts as a test particle and does not perturb, in any way, the circular motion of the primary bodies.

We adopt a special system of units in which the gravitational constant $G$, the sum of the masses of the primaries, the speed of light $c$ as well as the distance $R$ between the primaries are equal to unity. For the description of the planar motion of the test particle we choose a rotating reference frame, where the center of mass of the primaries coincides with its origin. The dimensionless masses of the primary bodies $P_1$ and $P_2$ are $m_1 = 1 - \mu$ and $m_2 = \mu$, respectively, where $\mu = m_2/(m_1 + m_2) \leqslant 1/2$ is the mass ratio. Moreover, the centers of both primaries are located on the $x$-axis and specifically at $(- \mu, 0)$ and $(1 - \mu, 0)$. In this article, we shall consider the case where the two primary bodies have equal masses (that is when $m_1 = m_2$, which is also known as the Copenhagen problem) and therefore $\mu = 1/2$.

According to \citet{DLCG17a} the time-independent effective potential function of the pseudo-Newtonian planar circular restricted three-body problem, with only the first correction terms, is
\begin{equation}
\Omega(x,y) = \frac{(1 - \mu)}{r_1} + \frac{\mu}{r_2} - \frac{\epsilon}{2c^4}\left(\frac{(1 - \mu)^3}{r_1^3} + \frac{\mu^3}{r_2^3}\right) + \frac{1}{2}\left(x^2 + y^2\right),
\label{eff}
\end{equation}
where of course $(x,y)$ are the coordinates of the test particle on the configuration plane, while
\begin{equation}
r_1 = \sqrt{\left(x + \mu\right)^2 + y^2}, \ \ \
r_2 = \sqrt{\left(x + \mu - 1\right)^2 + y^2},
\label{dist}
\end{equation}
are the distances of the test particle from the two primary bodies.

It is seen that the effective potential function (\ref{eff}) can be written as the sum of three terms
\begin{equation}
\Omega(x,y) = \Omega_{\rm CN} - \frac{\epsilon}{2c^4} \Omega _{\rm PN} + \frac{1}{2}\left(x^2 + y^2\right),
\label{eff2}
\end{equation}
where $\Omega_{\rm CN}$ are the terms of the classical Newtonian dynamics, while $\Omega_{\rm PN}$ contains the pseudo-Newtonian correction terms.

The dynamical quantity $\epsilon$ is a transition parameter with values in the interval $\epsilon \in [0, 1]$. When $\epsilon = 0$ we have the case of the classical Newtonian three-body problem, while when $\epsilon = 1$ we have the case of the full pseudo-Newtonian three-body problem.

The equations of motion of the test particle in the co-rotating reference frame read
\begin{equation}
\ddot{x} = \frac{\partial \Omega}{\partial x} + 2 \dot{y}, \ \ \
\ddot{y} = \frac{\partial \Omega}{\partial y} - 2 \dot{x},
\label{eqmot}
\end{equation}
where
\begin{align}
\Omega_x(x,y) &= \frac{\partial \Omega}{\partial x} = - \frac{m_1 x_1}{r_1^3} - \frac{m_2 x_2}{r_2^3}
+ \frac{3\epsilon}{2c^4}\left(\frac{m_1^3 x_1}{r_1^5} + \frac{m_2^3 x_2}{r_2^5}\right) + x, \\
\Omega_y(x,y) &= \frac{\partial \Omega}{\partial y} = - \frac{m_1 y}{r_1^3} - \frac{m_2 y}{r_2^3}
+ \frac{3\epsilon y}{2c^4}\left(\frac{m_1^3}{r_1^5} + \frac{m_2^3}{r_2^5}\right) + y,
\label{drv1}
\end{align}
with
$x_1 = x + \mu$ and $x_2 = x_1 - 1$. Similarly, the partial derivatives of the second order, which will be needed later for the multivariate Newton-Raphson iterative scheme, read
\begin{align}
\Omega_{xx}(x,y) &= \frac{\partial^2 \Omega}{\partial x^2} = - \frac{m_1\left(r_1^2 - 3x_1^2\right)}{r_1^5} - \frac{m_2\left(r_2^2 - 3x_2^2\right)}{r_2^5} \nonumber\\
&+ \frac{3\epsilon}{2c^4}\left(\frac{m_1^3\left(r_1^2 - 5x_1^2\right)}{r_1^7} + \frac{m_2^3\left(r_2^2 - 5x_2^2\right)}{r_2^7}\right) + 1, \\
\Omega_{xy}(x,y) &= \frac{\partial^2 \Omega}{\partial x \partial y} = 3\left(\frac{m_1 x_1}{r_1^5} + \frac{m_2 x_2}{r_2^5}\right)y \nonumber\\
&- \frac{15\epsilon y}{2c^4}\left(\frac{m_1^3 x_1}{r_1^7} + \frac{m_2^3 x_2}{r_2^7}\right), \\
\Omega_{yx}(x,y) &= \frac{\partial^2 \Omega}{\partial y \partial x} = \Omega_{xy}(x,y), \\
\Omega_{yy}(x,y) &= \frac{\partial^2 \Omega}{\partial y^2} = - \frac{m_1\left(r_1^2 - 3y^2\right)}{r_1^5} - \frac{m_2\left(r_2^2 - 3y^2\right)}{r_2^5} \nonumber\\
&+ \frac{3\epsilon}{2c^4}\left(\frac{m_1^3\left(r_1^2 - 5y^2\right)}{r_1^7} + \frac{m_2^3\left(r_2^2 - 5y^2\right)}{r_2^7}\right) + 1.
\label{drv2}
\end{align}

For the system of the differential equations (\ref{eqmot}) there is only one integral of motion (also known as the Jacobi integral) which is given by the following Hamiltonian
\begin{equation}
J(x,y,\dot{x},\dot{y}) = 2\Omega(x,y) - \left(\dot{x}^2 + \dot{y}^2 \right) = C,
\label{ham}
\end{equation}
where $\dot{x}$ and $\dot{y}$ are the velocities, while $C$ is the numerical value of the Jacobi constant which is conserved.

\section{Parametric evolution and stability of the equilibrium points}
\label{param}

For the existence of equilibrium points the necessary and sufficient conditions which must be fulfilled are
\begin{equation}
\dot{x} = \dot{y} = \ddot{x} = \ddot{y} = 0.
\label{lps0}
\end{equation}
For determining the coordinates $(x,y)$ of the coplanar equilibrium points we have to numerically solve the following system of partial differential equations
\begin{equation}
\Omega_x(x,y) = 0, \ \ \ \Omega_y(x,y) = 0.
\label{lps}
\end{equation}

\begin{figure*}[!t]
\centering
\resizebox{\hsize}{!}{\includegraphics{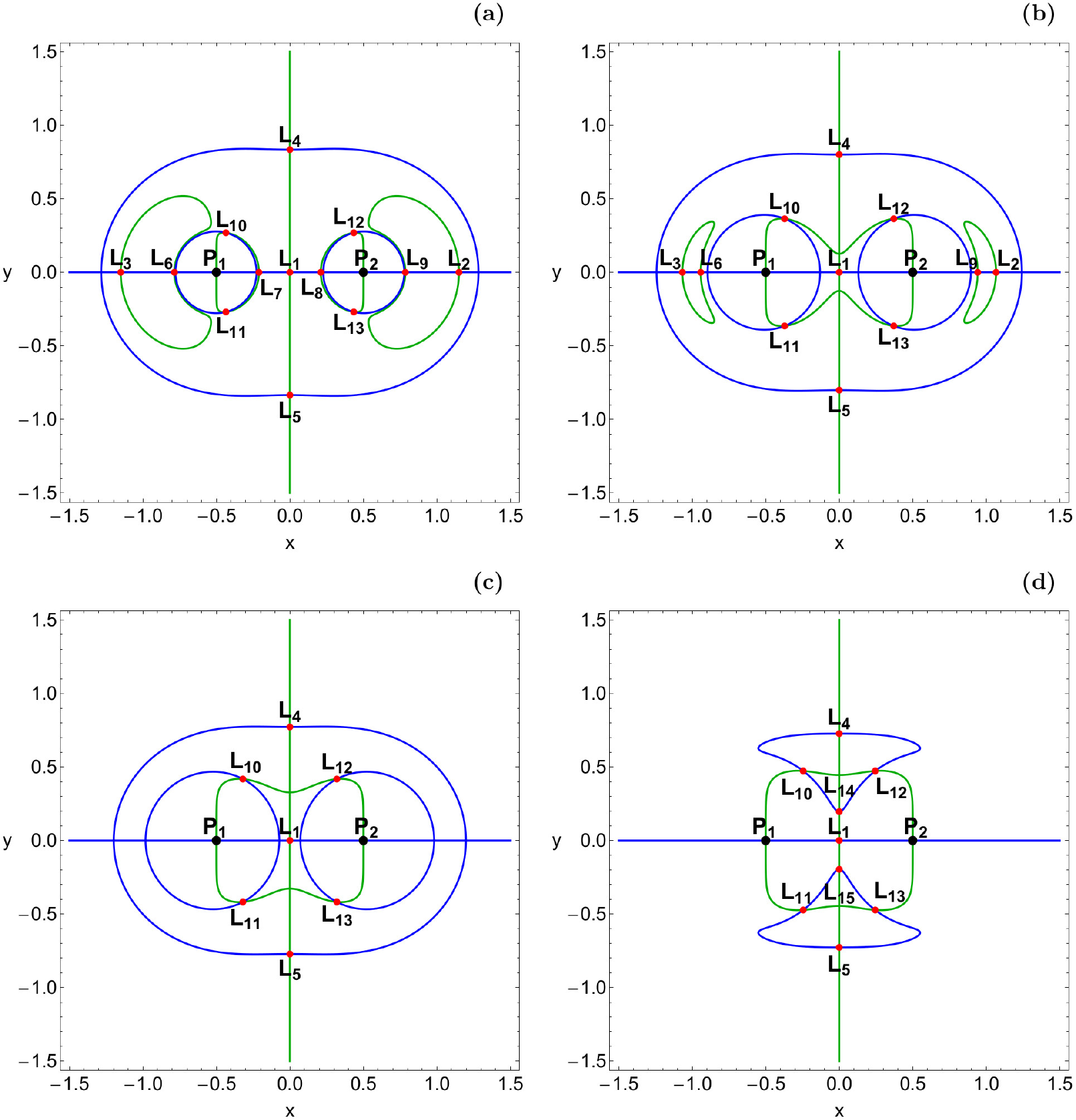}}
\caption{Locations of the positions (red dots) and numbering of the equilibrium points $(L_i, \ i=1,15)$ through the intersections of $\Omega_x = 0$ (green) and $\Omega_y = 0$ (blue), when (a-upper left): $\epsilon = 0.2$, (b-upper right): $\epsilon = 0.375$, (c-lower left): $\epsilon = 0.5$, and (d-lower right): $\epsilon = 0.65$. The black dots denote the centers $(P_i, \ i=1,2)$ of the two primaries.}
\label{lgs}
\end{figure*}

\begin{figure*}[!t]
\centering
\resizebox{\hsize}{!}{\includegraphics{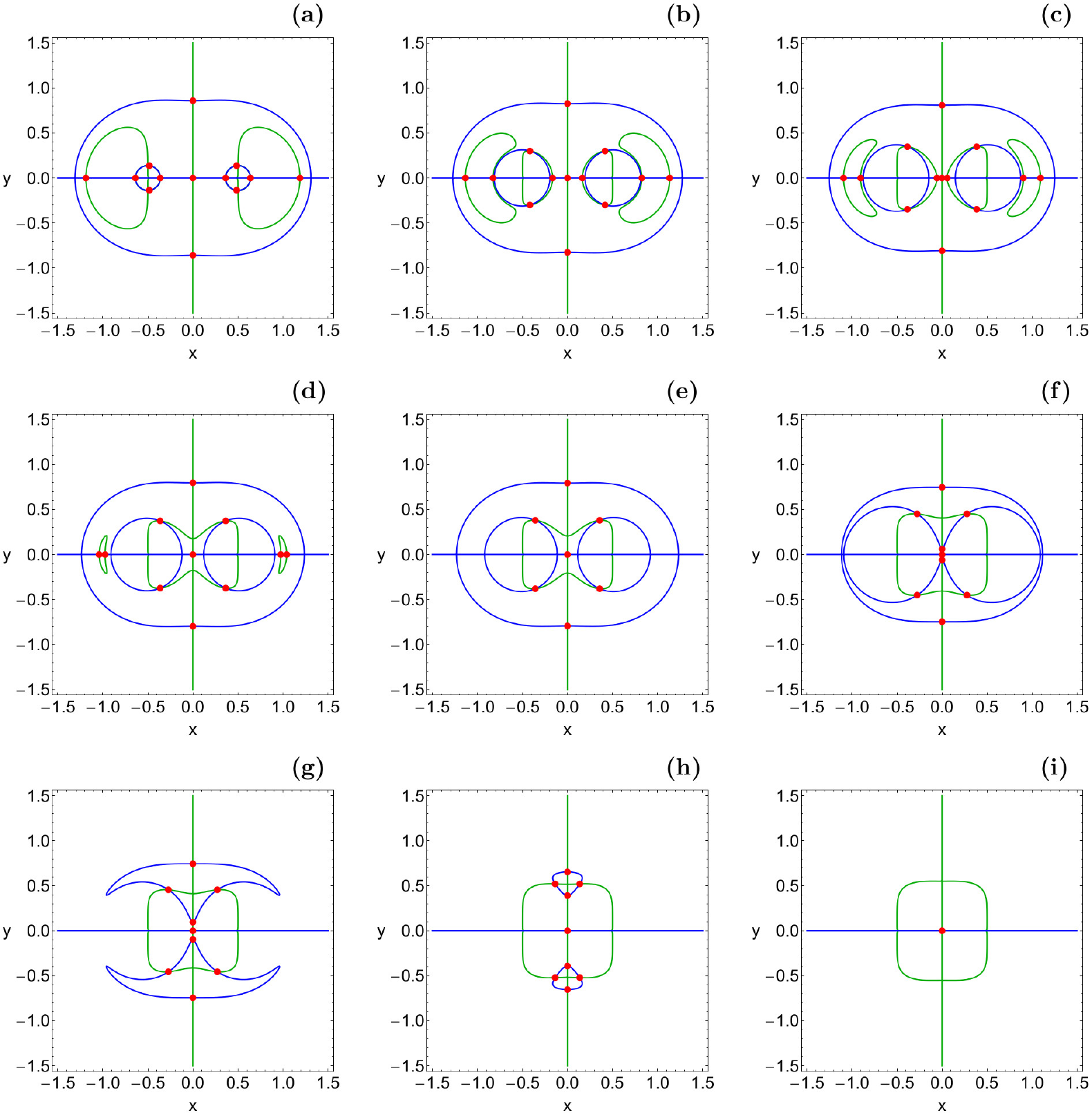}}
\caption{Evolution of the locations of the positions (red dots) of the equilibrium points $(L_i, \ i=1,15)$ when the transition parameter varies in the interval $(0,1]$. In particular (a): $\epsilon = 0.05$, (b): $\epsilon = 0.25$, (c): $\epsilon = 0.34$, (d): $\epsilon = 0.395$, (e): $\epsilon = 0.41$, (f): $\epsilon = 0.59$, (g): $\epsilon = 0.60$, (h): $\epsilon = 0.80$, (i): $\epsilon = 0.90$.}
\label{conts}
\end{figure*}

The total number of the equilibrium points in the pseudo-Newtonian planar circular restricted three-body problem, with two equal masses, is not constant but it strongly depends on the value of the transition parameter $\epsilon$. More precisely
\begin{itemize}
  \item When $\epsilon = 0$ we have the case of the classical three-body problem, so there are the usual five equilibrium points, three collinear, ($L_1$, $L_2$, and $L_3$), and two triangular ($L_4$ and $L_5$).
  \item When $\epsilon \in (0, 0.35416667]$ there exist thirteen equilibrium points (see panel (a) of Fig. \ref{lgs}). On the $x$ axis four additional collinear points ($L_6$, $L_7$, $L_8$, and $L_9$) emerge, while four more points ($L_{10}$, $L_{11}$, $L_{12}$, and $L_{13}$) appear on the configuration $(x,y)$ plane.
  \item When $\epsilon \in [0.35416668, 0.40306154]$ there exist eleven equilibrium points (see panel (b) of Fig. \ref{lgs}). In this case the collinear points $L_7$ and $L_8$ are not present.
  \item When $\epsilon \in [0.40306155, 0.58333333]$ there exist seven equilibrium points (see panel (c) of Fig. \ref{lgs}). In this case the collinear points $L_2$, $L_3$, $L_6$, and $L_9$ disappear.
  \item When $\epsilon \in [0.58333334, 0.86861363]$ there exist nine equilibrium points (see panel (d) of Fig. \ref{lgs}). Two new equilibrium points, $L_{14}$ and $L_{15}$, emerge on the vertical $y$ axis.
  \item When $\epsilon \in [0.86861364, 1]$ only the libration point $L_1$, located at the origin $(0,0)$, survives.
\end{itemize}
The values $\epsilon_1 = 0.35416667$, $\epsilon_2 = 0.40306154$, $\epsilon_3 = 0.58333333$, and $\epsilon_4 = 0.86861363$ are critical values of the transition parameter, since they delimit the ranges of several intervals with different number of equilibrium points.

\begin{figure*}[!t]
\centering
\resizebox{\hsize}{!}{\includegraphics{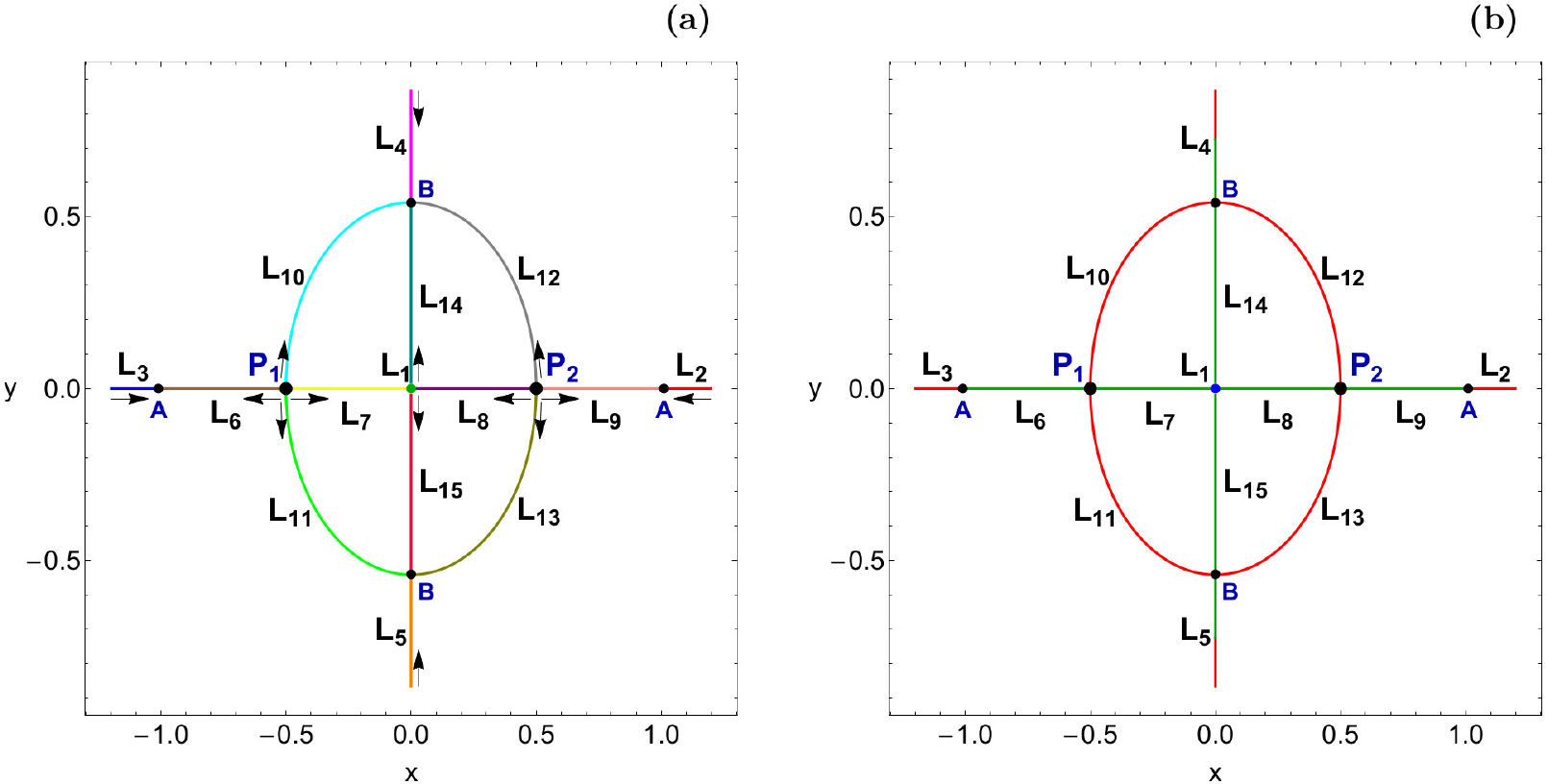}}
\caption{The parametric evolution of (a-left): the position and (b-right): the stability (green) or instability (red) of the equilibrium points in the pseudo-Newtonian planar circular restricted three-body problem with equal masses, when $\epsilon \in (0, 1]$. The arrows indicate the movement direction of the equilibrium points as the value of the transition parameter $\epsilon$ increases. The big black dots pinpoint the fixed centers of the two primaries, while the small black dots (points A and B) correspond to the critical values $\epsilon_2$ and $\epsilon_4$, respectively.}
\label{evol}
\end{figure*}

The position of all the equilibrium points is defined through the intersections of the equations $\Omega_x = 0, \ \Omega_y = 0$. In Fig. \ref{lgs}(a-d) we see how the intersections of the first order partial derivatives determine, in each case, the position of the libration points when (a): $\epsilon = 0.2$, (b): $\epsilon = 0.373$, (c): $\epsilon = 0.5$, (d): $\epsilon = 0.65$. In the same figure we provide the numbering of all the libration points $L_i, \ i=1,15$. Furthermore, in Fig. \ref{conts} we present how the number and the exact positions of the equilibrium points evolve as the value of the transition parameter varies in the interval $(0,1]$.

It would be very interesting to obtain the exact evolution of the positions of the libration points as a function of the transition parameter $\epsilon$, when $\epsilon \in (0,1]$. Our numerical analysis is illustrated in Fig. \ref{evol}, where the parametric evolution of all the equilibrium points, on the configuration $(x,y)$ plane, is presented. One may observe that as soon as $\epsilon$ is just above zero, eight equilibrium points (two sets of four) emerge from the centers $P_1$ and $P_2$. As the value of $\epsilon$ grows the collinear equilibrium points $L_i, \ i = 6,...,9$ move away from the centers. In particular, $L_6$ and $L_9$ move towards $L_3$ and $L_2$, respectively, while on the other hand $L_7$ and $L_8$ move towards the center. When $\epsilon = \epsilon_1$ the equilibrium points $L_7$ and $L_8$ collide with the central libration point $L_1$ and they disappear. In the same vein when $\epsilon = \epsilon_2$ $L_6$ collides with $L_3$ and at the same time $L_9$ collides with $L_2$ thus annihilating each other. When $\epsilon = \epsilon_3$ the phenomenon of the creation of new equilibrium points occurs, since two new libration points $L_{14}$ and $L_{15}$ emerge from the center. As soon as $\epsilon > \epsilon_3$ these new points move on the vertical $y$ axis and away from $L_1$. It is seen that $L_4$, $L_{10}$, $L_{12}$, and $L_{14}$ (the same applies for $L_5$, $L_{11}$, $L_{13}$, and $L_{15}$) move on a collision course. The collision occurs when $\epsilon = \epsilon_4$ and all these libration points are being destroyed in two sets. Finally when $\epsilon > \epsilon_4$ only the central equilibrium point $L_1$ survives and remains present until $\epsilon = 1$. At this point it should be emphasized that the centers of the two primaries $P_1$ and $P_2$ are completely unaffected by the shift of the transition parameter.

In order to determine the linear stability of an equilibrium point the origin of the reference frame must be transferred at the exact position $(x_0,y_0)$ of the libration point through the transformation
\begin{equation}
x = x_0 + \xi, \ \ \ y = y_0 + \eta.
\label{trans}
\end{equation}
The next step is to expand the system of the equations of motion (\ref{eqmot}) into first-order terms, with respect to $\xi$ and $\eta$.
\begin{equation}
\dot{{\bf{\Xi}}} = A {\bf{\Xi}}, \ \ {\bf{\Xi}} = \left(\xi, \eta, \dot{\xi}, \dot{\eta}\right)^{\rm T},
\label{ls}
\end{equation}
where ${\bf{\Xi}}$ is the state vector of the test particle with respect to the equilibrium points, while $A$ is the time-independent coefficient matrix of variations
\begin{equation}
A =
\begin{bmatrix}
    0 & 0 & 1 & 0 \\
    0 & 0 & 0 & 1 \\
    \Omega_{xx}^0 & \Omega_{xy}^0 &  0 & 2 \\
    \Omega_{yx}^0 & \Omega_{yy}^0 & -2 & 0
\end{bmatrix},
\end{equation}
where the superscript 0, at the partial derivatives of second order, denotes evaluation at the position of the equilibrium point $(x_0, y_0)$. The new linearized system describes infinitesimal motions near an equilibrium point.

The characteristic equation of the linear system (\ref{ls}) is quadratic with respect to $\Lambda = \lambda^2$ and it is given by
\begin{equation}
\alpha \Lambda^2 + b \Lambda + c = 0,
\label{ceq}
\end{equation}
where
\begin{equation}
\alpha = 1, \ \ \ b = 4 - \Omega_{xx}^0 - \Omega_{yy}^0, \ \ \ c = \Omega_{xx}^0 \Omega_{yy}^0 - \Omega_{xy}^0 \Omega_{yx}^0.
\end{equation}

The necessary and sufficient condition for an equilibrium point to be stable is all roots of the characteristic equation to be pure imaginary. This means that the following three conditions must be simultaneously fulfilled
\begin{equation}
b > 0, \ \ \ c > 0, \ \ \ D = b^2 - 4 a c > 0.
\end{equation}
This fact ensures that the characteristic equation (\ref{ceq}) has two real negative roots $\Lambda_{1,2}$, which consequently implies that there are four pure imaginary roots for $\lambda$.

Since we already know the exact positions $(x_0,y_0)$ of the libration points, we can insert them into the characteristic equation (\ref{ceq}) and therefore determine the stability of the equilibrium points, through the nature of the four roots. Our numerical analysis suggests that most of the equilibrium points are either stable or unstable when the transition parameter $\epsilon$ varies in the interval $(0,1]$. In particular, $L_2$, $L_3$, $L_{10}$, $L_{11}$, $L_{12}$, and $L_{13}$ are always unstable, while $L_6$, $L_7$, $L_8$, $L_9$, $L_{14}$, and $L_{15}$ are always stable. The equilibrium points $L_1$, $L_4$, and $L_5$ on the other hand, can be either stable or unstable, depending of course on the particular value of $\epsilon$. In panel (b) of Fig. \ref{evol} we illustrate the evolution of the stability of all the equilibrium points, when $\epsilon \in (0,1]$. Our numerical computations suggest that the central libration point $L_1$ is stable only when $\epsilon \in [\epsilon_1, \epsilon_3]$, while the triangular points $L_4$ and $L_5$ are stable only when $\epsilon$ lies in the interval $[0.65712024,\epsilon_4]$.

\section{The basins of attraction}
\label{bas}

Over the years many methods, for solving numerically systems of non-linear equations, have been developed. Perhaps the most well-known method of all is the Newton-Raphson method. A system of multivariate functions $f({\bf{x}}) = 0$ can be solved using the following iterative scheme
\begin{equation}
{\bf{x}}_{n+1} = {\bf{x}}_{n} - J^{-1}f({\bf{x}}_{n}),
\label{sch}
\end{equation}
where $f({\bf{x_n}})$ is the system of equations, while $J^{-1}$ is the corresponding inverse Jacobian matrix. In our case the system of differential equations is described in Eqs. (\ref{lps}).

The iterative formulae for each coordinate $(x,y)$, derived from scheme (\ref{sch}), are
\begin{align}
x_{n+1} &= x_n - \left( \frac{\Omega_x \Omega_{yy} - \Omega_y \Omega_{xy}}{\Omega_{yy} \Omega_{xx} - \Omega^2_{xy}} \right)_{(x_n,y_n)}, \nonumber\\
y_{n+1} &= y_n + \left( \frac{\Omega_x \Omega_{yx} - \Omega_y \Omega_{xx}}{\Omega_{yy} \Omega_{xx} - \Omega^2_{xy}} \right)_{(x_n,y_n)},
\label{nrm}
\end{align}
where $x_n$, $y_n$ are the values of the $x$ and $y$ coordinates at the $n$-th step of the iterative process.

\begin{figure*}[!t]
\centering
\resizebox{\hsize}{!}{\includegraphics{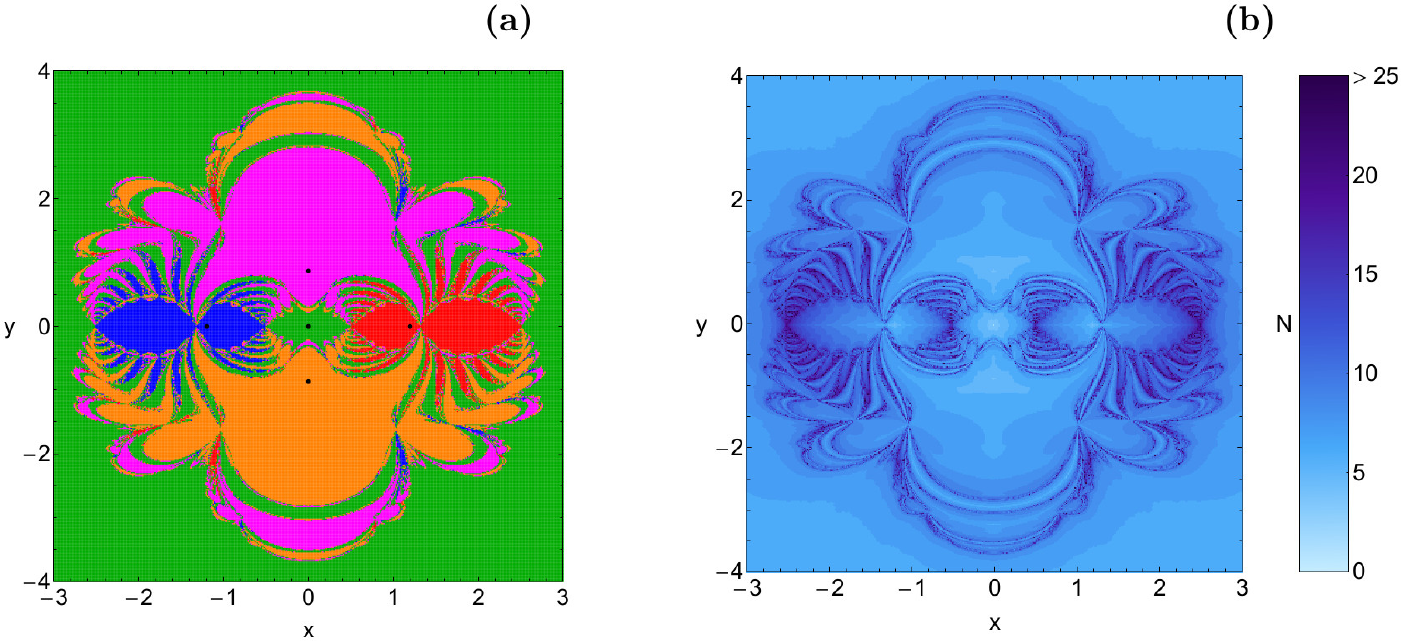}}
\caption{(a-left): The Newton-Raphson basins of attraction on the configuration $(x,y)$ plane for the classical Newtonian case, where $\epsilon = 0$. The positions of the five equilibrium points are indicated by black dots. The color code denoting the five attractors (equilibrium points) is as follows: $L_1$ (green); $L_2$ (red); $L_3$ (blue); $L_4$ (magenta); $L_5$ (orange); non-converging points (white). (b-right): The distribution of the corresponding number $N$ of required iterations for obtaining the Newton-Raphson basins of attraction shown in panel (a).}
\label{cn}
\end{figure*}

The numerical algorithm of the Newton-Raphson method works as follows: The code is activated when an initial condition $(x_0,y_0)$ on the configuration plane is inserted, while the iterative procedure continues until an attractor of the system is reached, with the desired accuracy. If the iterative procedure leads to one of the attractors then we say that the method converges for the particular initial condition. However, in general terms, not all initial conditions converges to an attractor of the system. All the initial conditions that lead to a specific final state (attractor) compose the Newton-Raphson basins of attraction, which are also known as basins of convergence or even as attracting regions/domains. At this point it should be highly noticed that the Newton-Raphson basins of attraction should not be mistaken, by no means, with the classical basins of attraction which exist in the case of dissipative systems. The Newton-Raphson basins of attraction are just a numerical artifact produced by an iterative scheme, while on the other hand the basins of attraction in dissipative systems correspond to a real observed phenomenon (attraction).

Nevertheless, the determination of the Newton-Raphson basins of attraction is very important because they reflect some of the most intrinsic qualitative properties of the dynamical system. This is true because the iterative formulae of Eqs. (\ref{nrm}) contain both the first and second order derivatives of the effective potential function $\Omega(x,y)$.

In order to unveil the basins of convergence we have to perform a double scan of the configuration $(x,y)$ plane. For this purpose we define dense uniform grids of $1024 \times 1024$ $(x_0,y_0)$ nodes which shall be used as initial conditions of the numerical algorithm. Of course the initial conditions corresponding to the centers $P_1$ and $P_2$ of the two primaries are excluded from all grids, because for these initial conditions the distances $r_i$, $i = 1,2$ to the primaries are equal to zero and consequently several terms, entering formulae (\ref{nrm}), become singular. During the classification of the initial conditions we also keep records of the number $N$ of iterations, required for the desired accuracy. Obviously, the better the desired accuracy the higher the required iterations. In our calculations the maximum number of iterations is set to $N_{\rm max} = 500$, while the iterative procedure stops only when an accuracy of $10^{-15}$ is reached, regarding the position of the attractors.

\begin{figure*}[!t]
\centering
\resizebox{0.70\hsize}{!}{\includegraphics{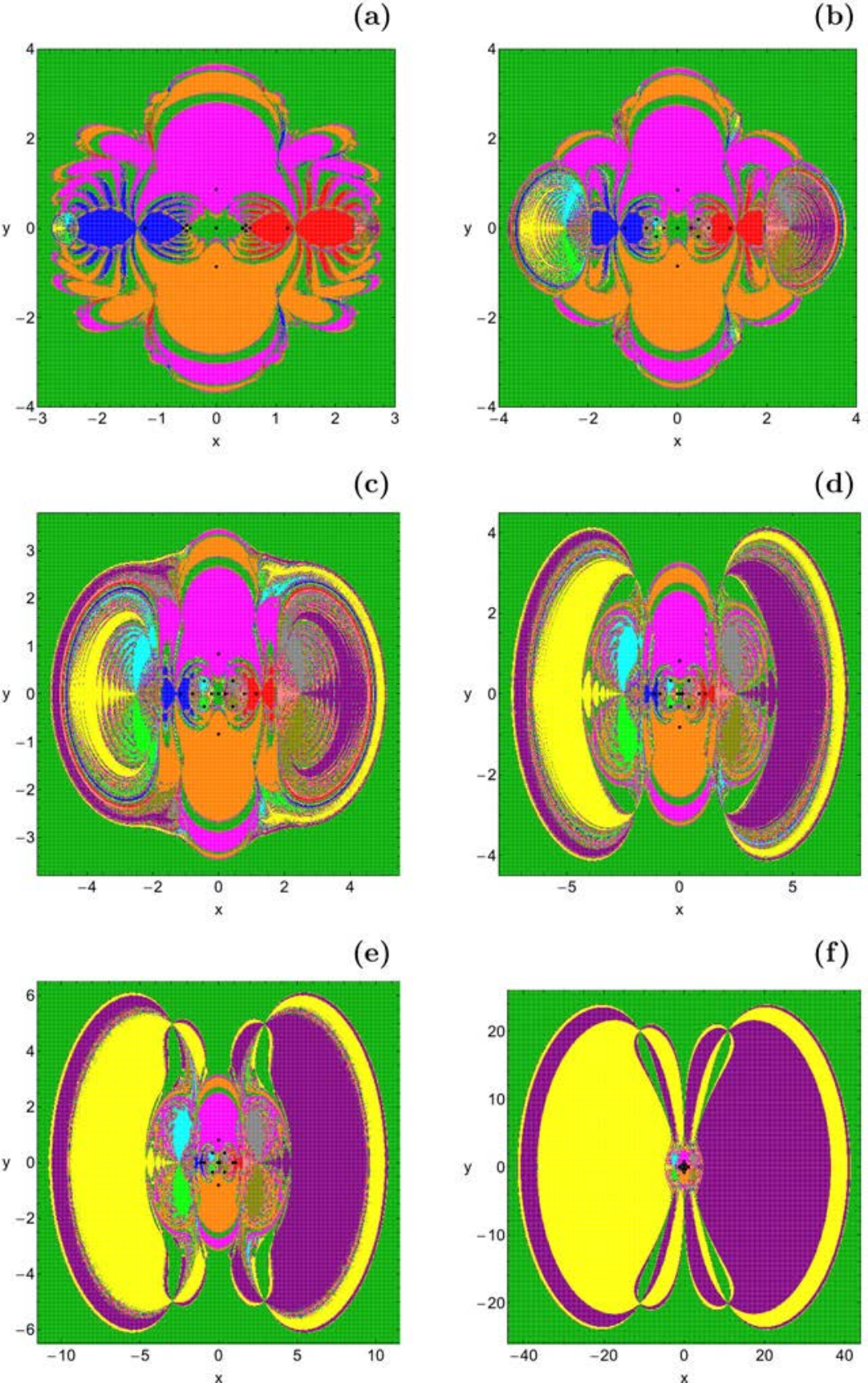}}
\caption{The Newton-Raphson basins of attraction on the configuration $(x,y)$ plane for the first case, where thirteen equilibrium points are present. (a): $\epsilon = 0.01$; (b): $\epsilon = 0.1$; (c): $\epsilon = 0.2$; (d): $\epsilon = 0.3$; (e): $m\epsilon = 0.34$; (f): $\epsilon = 0.3541$. The positions of the equilibrium points are indicated by black dots. The color code, denoting the 13 attractors, is as follows: $L_1$ (green); $L_2$ (red); $L_3$ (blue); $L_4$ (magenta); $L_5$ (orange); $L_6$ (brown); $L_7$ (yellow); $L_8$ (purple); $L_9$ (pink); $L_{10}$ (cyan); $L_{11}$ (light green); $L_{12}$ (gray); $L_{13}$ (olive); non-converging points (white).}
\label{r1}
\end{figure*}

\begin{figure*}[!t]
\centering
\resizebox{0.80\hsize}{!}{\includegraphics{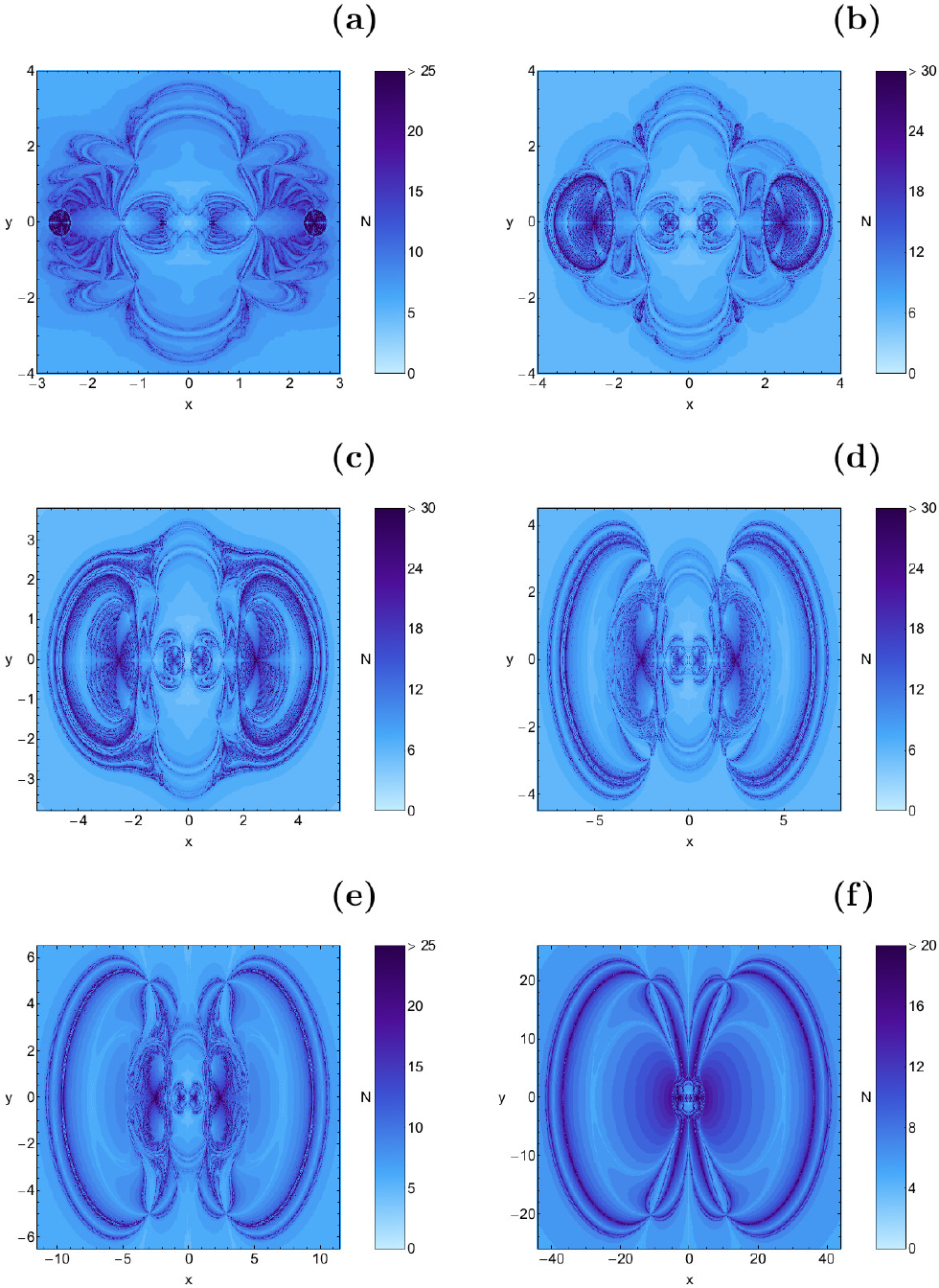}}
\caption{The distribution of the corresponding number $N$ of required iterations for obtaining the Newton-Raphson basins of attraction shown in Fig. \ref{r1}(a-f). The non-converging points are shown in white.}
\label{r1n}
\end{figure*}

\begin{figure*}[!t]
\centering
\resizebox{0.70\hsize}{!}{\includegraphics{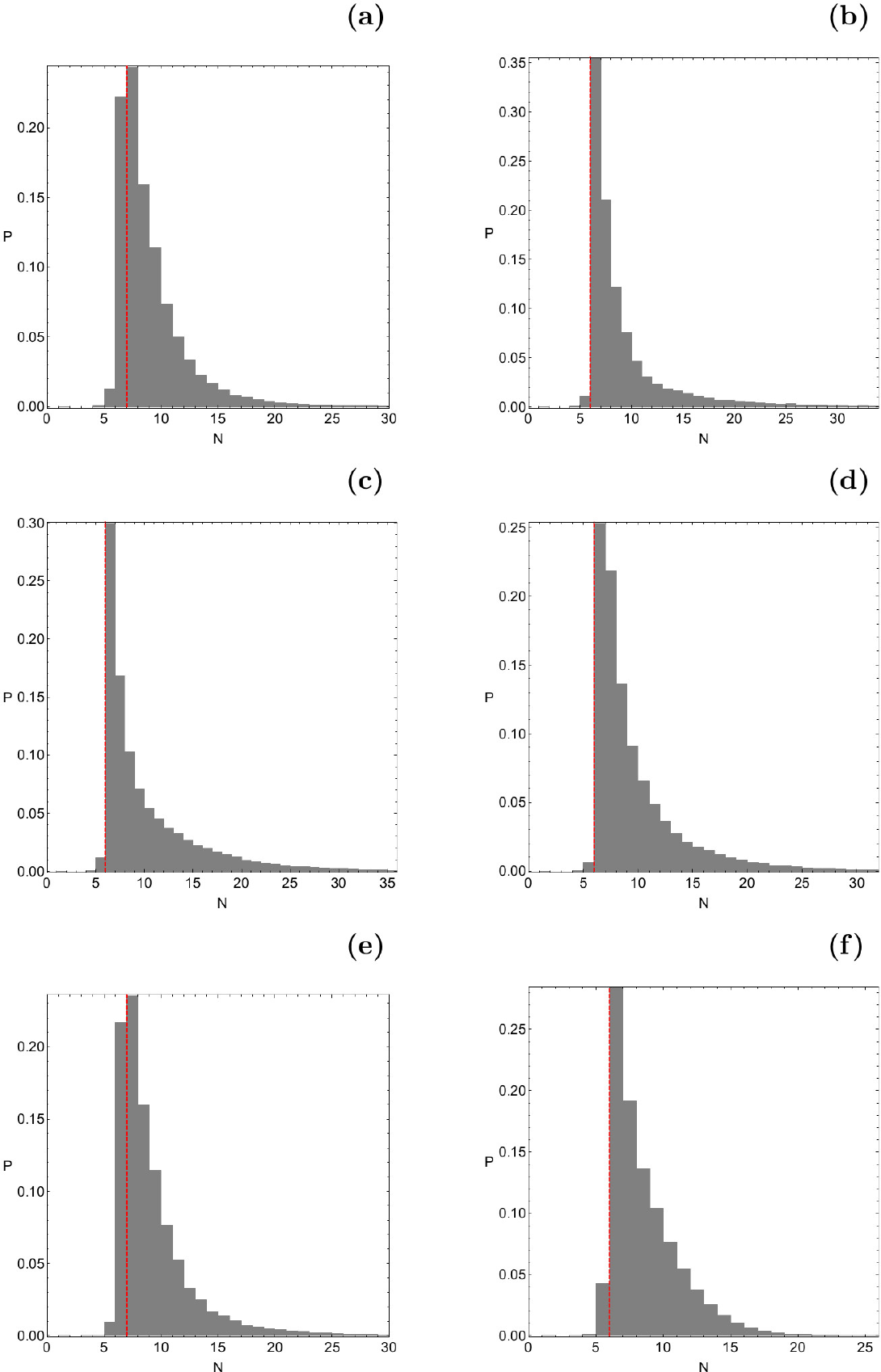}}
\caption{The corresponding probability distribution of required iterations for obtaining the Newton-Raphson basins of attraction shown in Fig. \ref{r1}(a-f). The vertical dashed red line indicates, in each case, the most probable number $N^{*}$ of iterations.}
\label{r1p}
\end{figure*}

\begin{figure*}[!t]
\centering
\resizebox{\hsize}{!}{\includegraphics{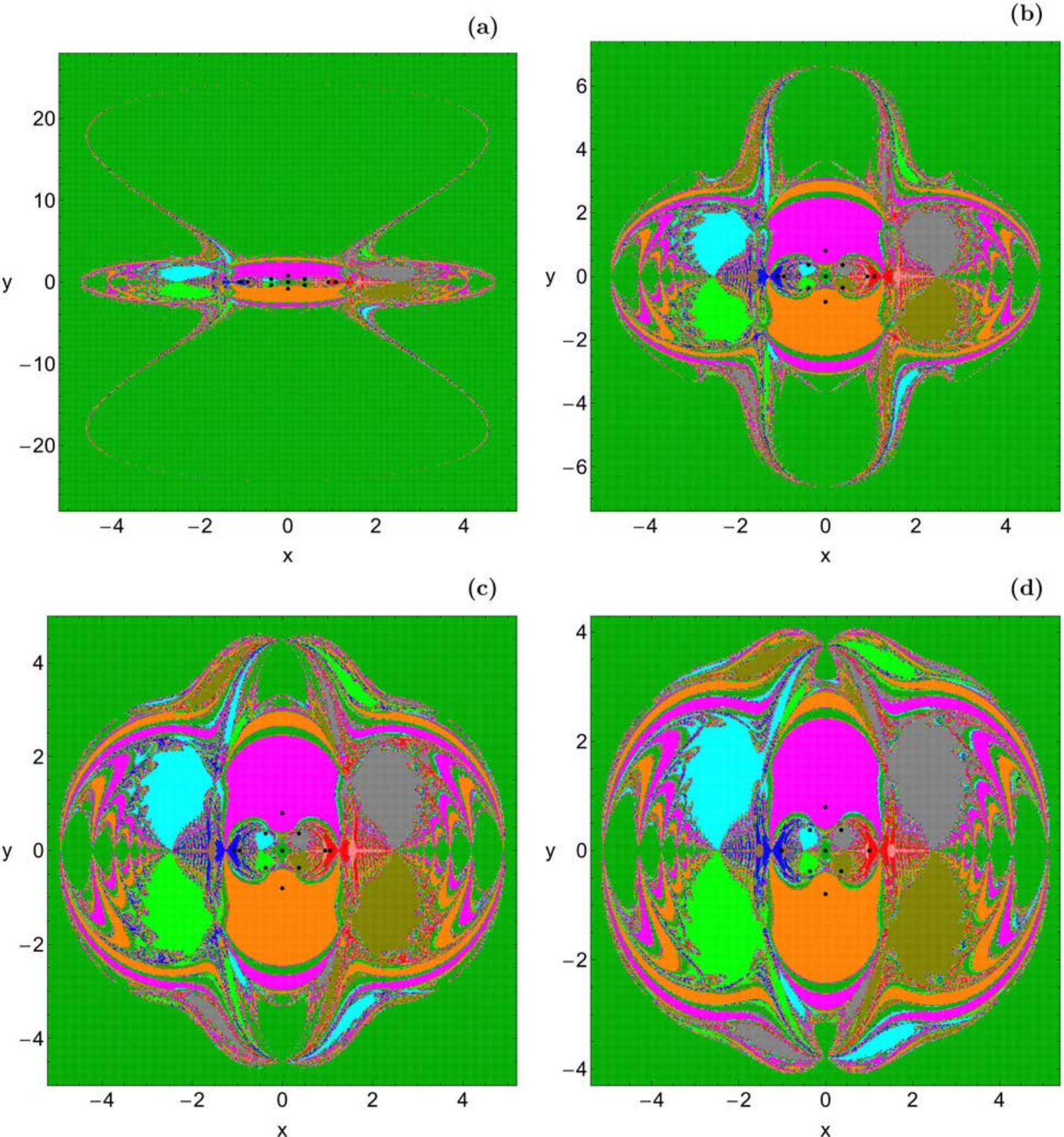}}
\caption{The Newton-Raphson basins of attraction on the configuration $(x,y)$ plane for the second case, where eleven equilibrium points are present. (a): $\epsilon = 0.3542$; (b): $\epsilon = 0.36$; (c): $\epsilon = 0.38$; (d): $\epsilon = 0.403$. The positions of the equilibrium points are indicated by black dots. The color code is the same as in Fig. \ref{r1}.}
\label{r2}
\end{figure*}

\begin{figure*}[!t]
\centering
\resizebox{\hsize}{!}{\includegraphics{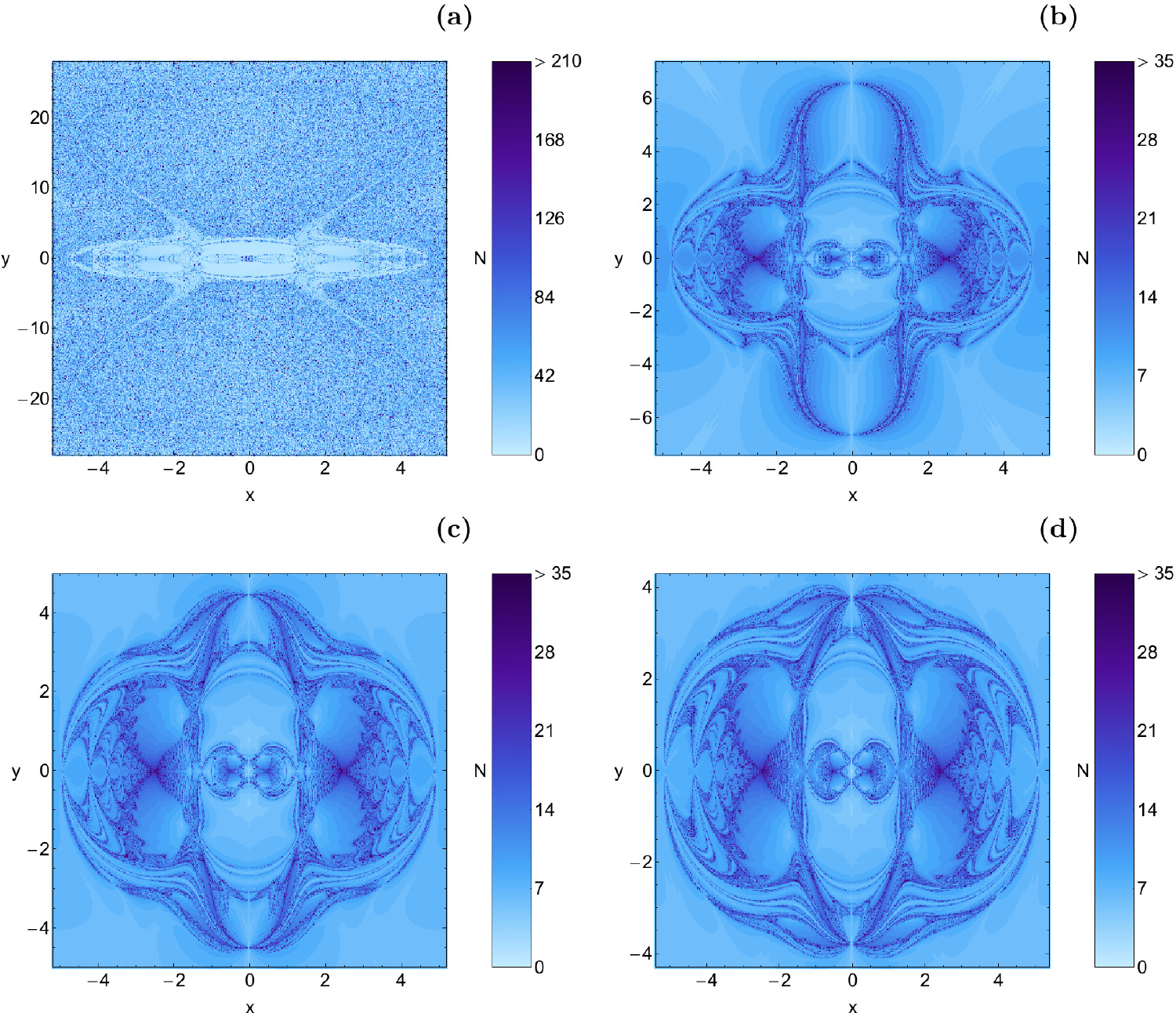}}
\caption{The distribution of the corresponding number $N$ of required iterations for obtaining the Newton-Raphson basins of attraction shown in Fig. \ref{r2}(a-d). The non-converging points are shown in white.}
\label{r2n}
\end{figure*}

\begin{figure*}[!t]
\centering
\resizebox{\hsize}{!}{\includegraphics{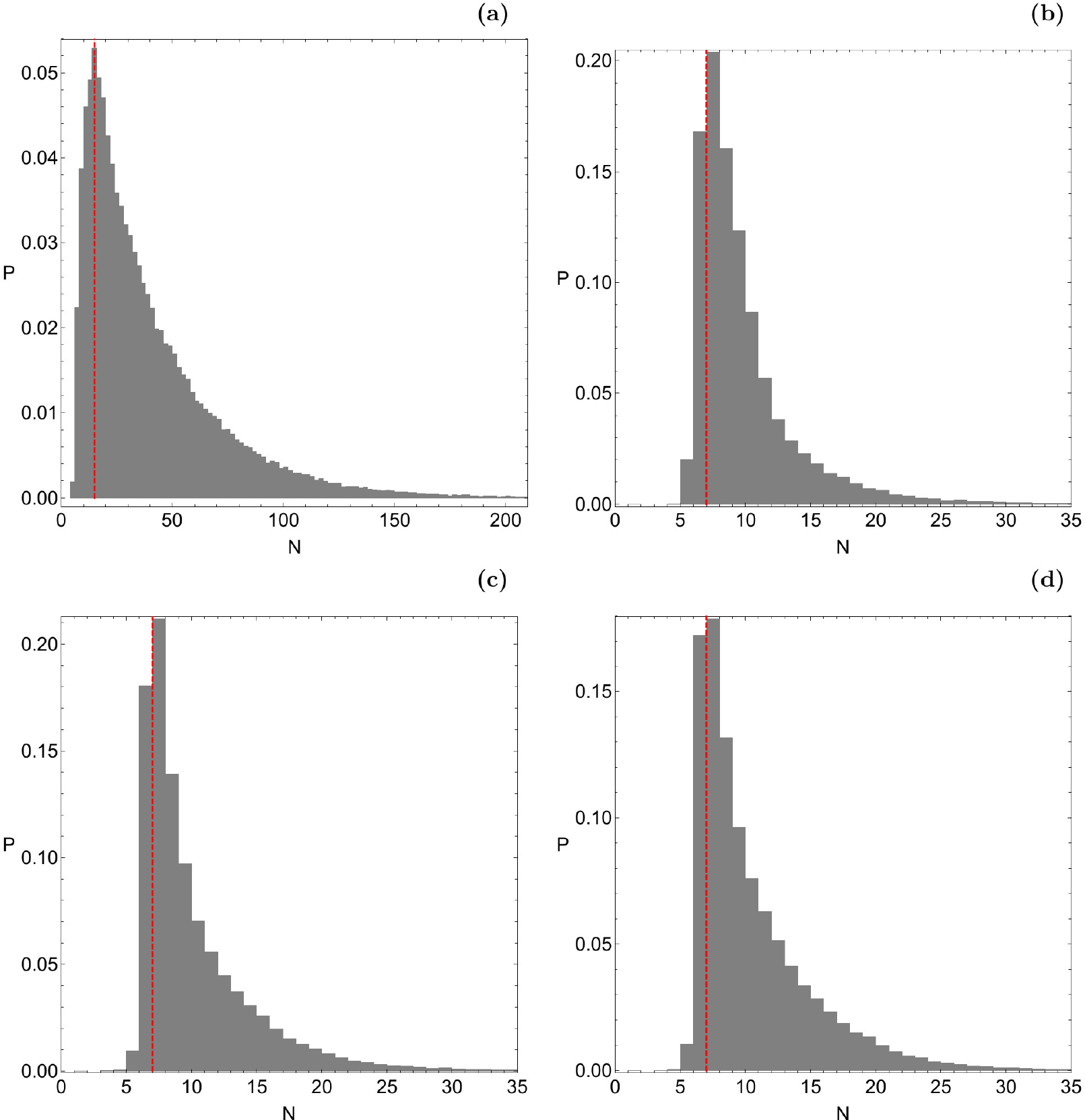}}
\caption{The corresponding probability distribution of required iterations for obtaining the Newton-Raphson basins of attraction shown in Fig. \ref{r2}(a-d). The vertical dashed red line indicates, in each case, the most probable number $N^{*}$ of iterations.}
\label{r2p}
\end{figure*}

\begin{figure*}[!t]
\centering
\resizebox{\hsize}{!}{\includegraphics{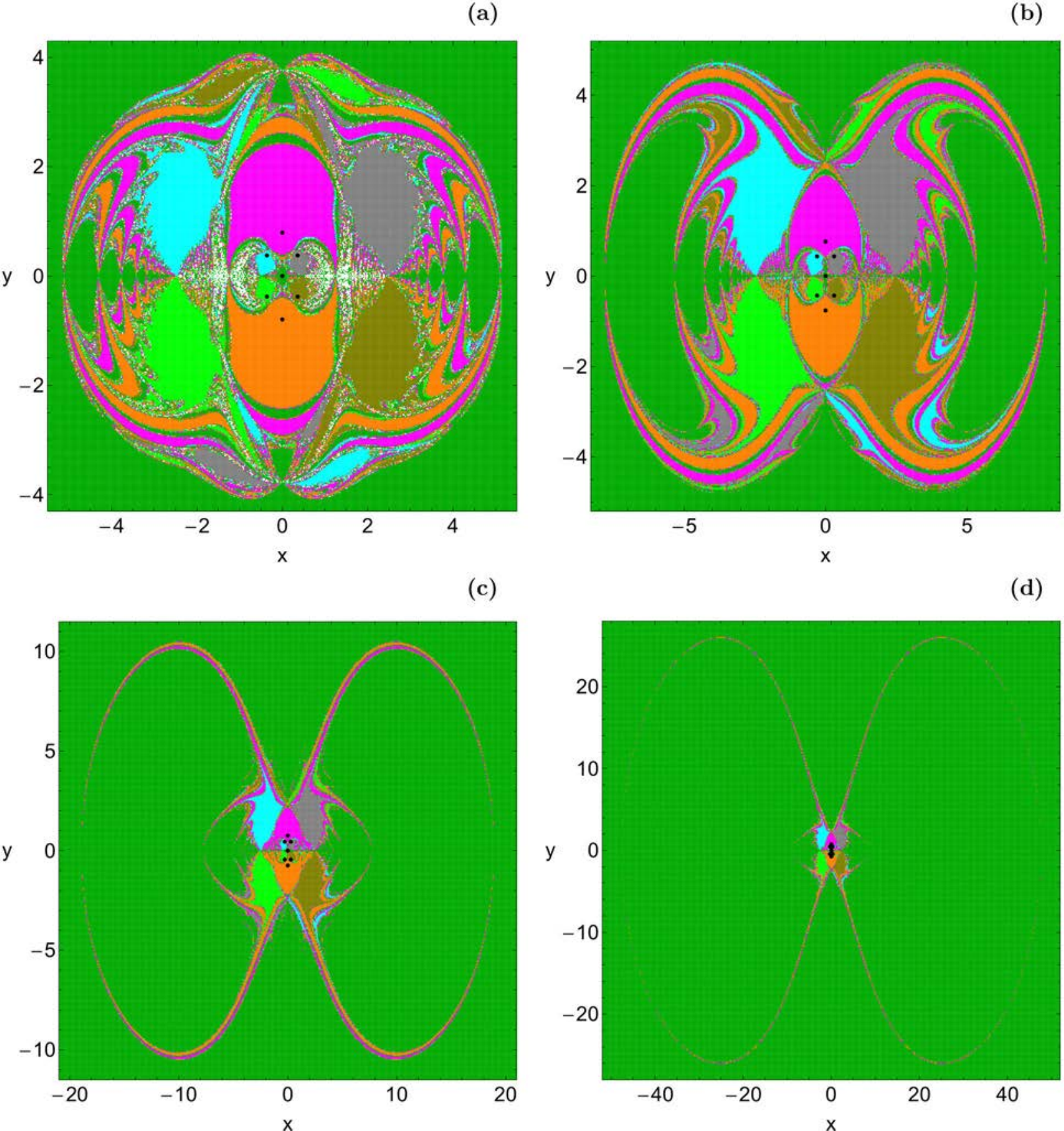}}
\caption{The Newton-Raphson basins of attraction on the configuration $(x,y)$ plane for the third case, where seven equilibrium points are present. (a): $\epsilon = 0.4031$; (b): $\epsilon = 0.54$; (c): $\epsilon = 0.582$; (d): $\epsilon = 0.5833$. The positions of the equilibrium points are indicated by black dots. The color code is the same as in Fig. \ref{r1}.}
\label{r3}
\end{figure*}

\begin{figure*}[!t]
\centering
\resizebox{\hsize}{!}{\includegraphics{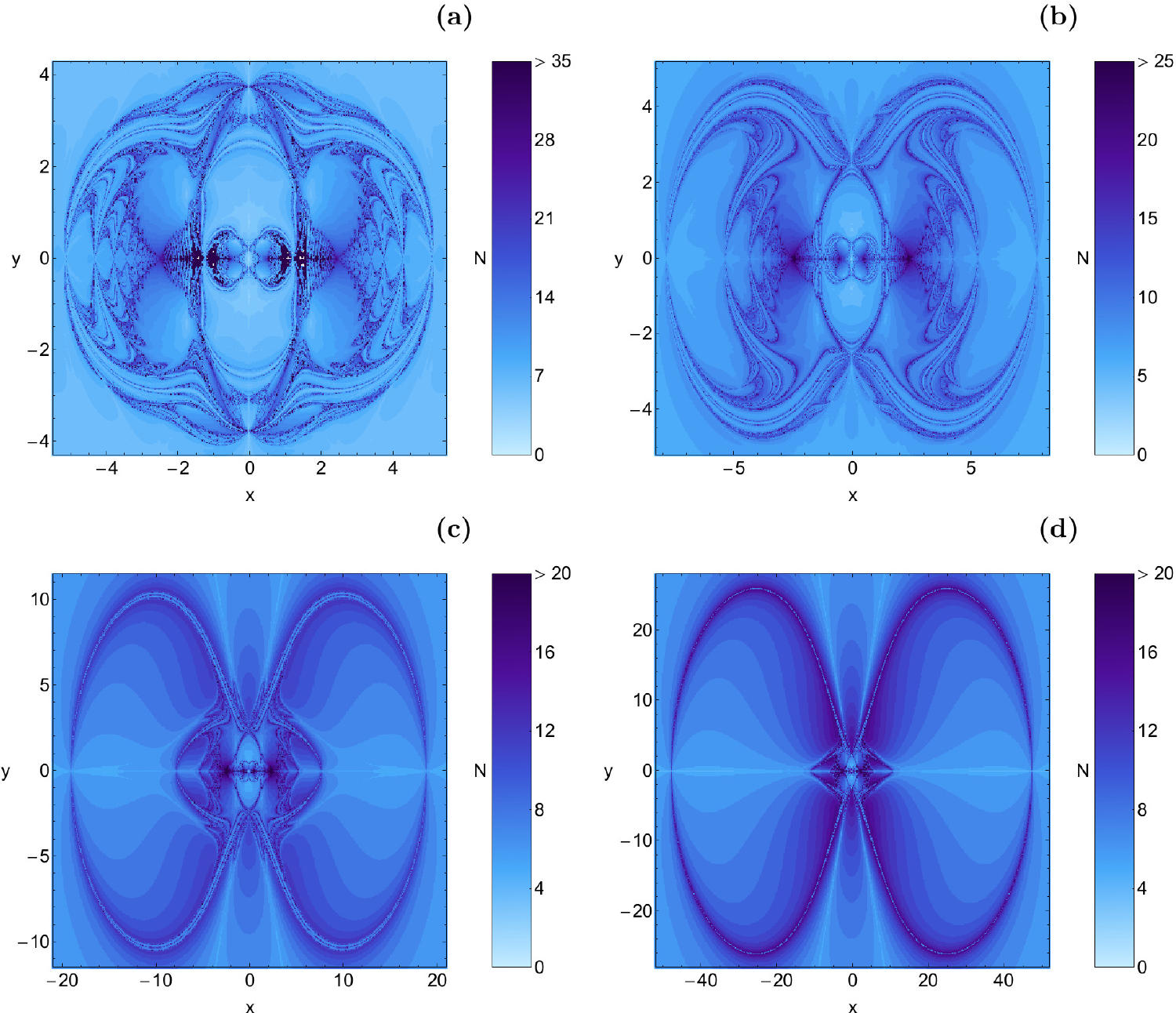}}
\caption{The distribution of the corresponding number $N$ of required iterations for obtaining the Newton-Raphson basins of attraction shown in Fig. \ref{r3}(a-d). The non-converging points are shown in white.}
\label{r3n}
\end{figure*}

\begin{figure*}[!t]
\centering
\resizebox{\hsize}{!}{\includegraphics{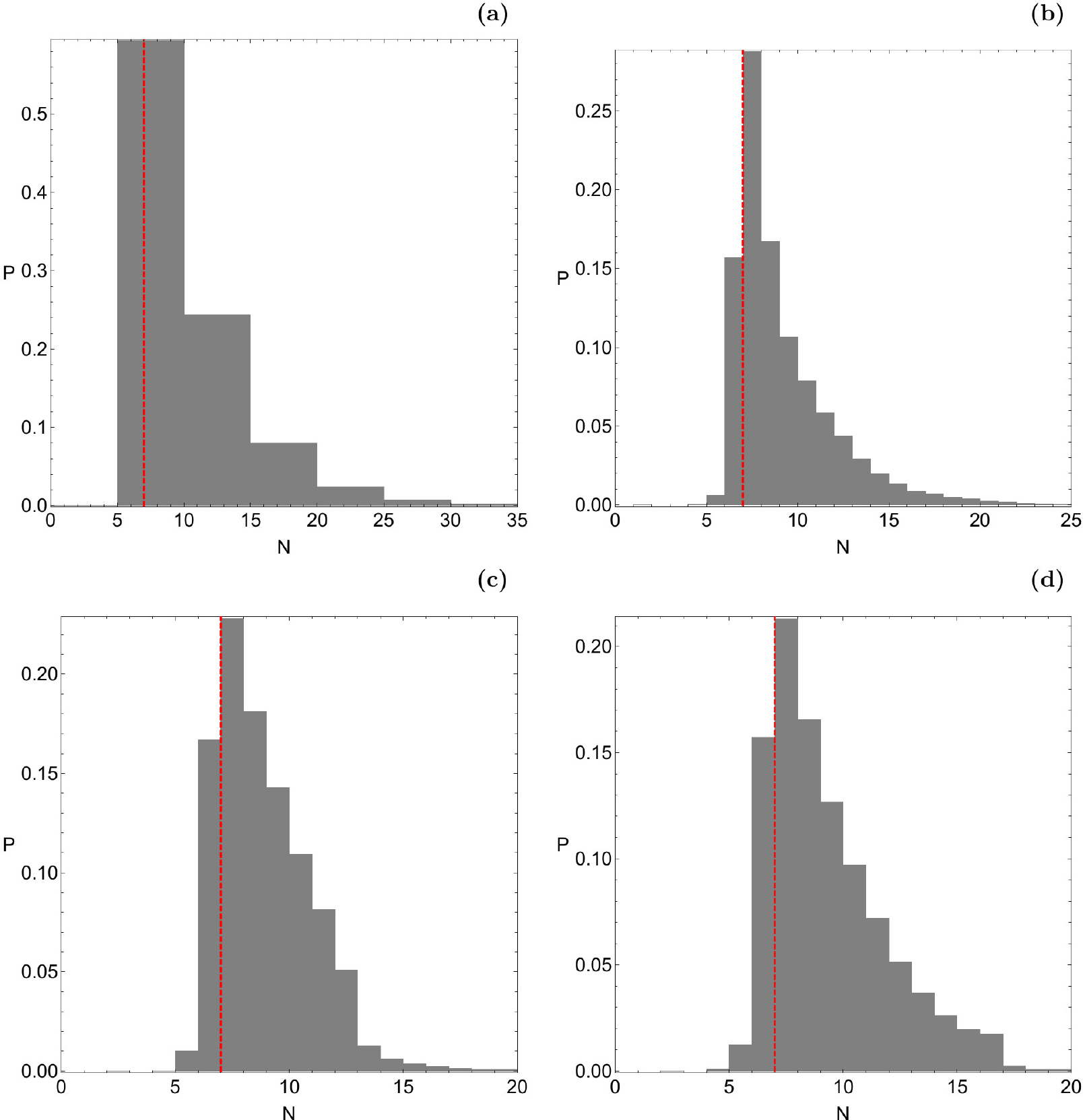}}
\caption{The corresponding probability distribution of required iterations for obtaining the Newton-Raphson basins of attraction shown in Fig. \ref{r3}(a-d). The vertical dashed red line indicates, in each case, the most probable number $N^{*}$ of iterations.}
\label{r3p}
\end{figure*}

\begin{figure*}[!t]
\centering
\resizebox{\hsize}{!}{\includegraphics{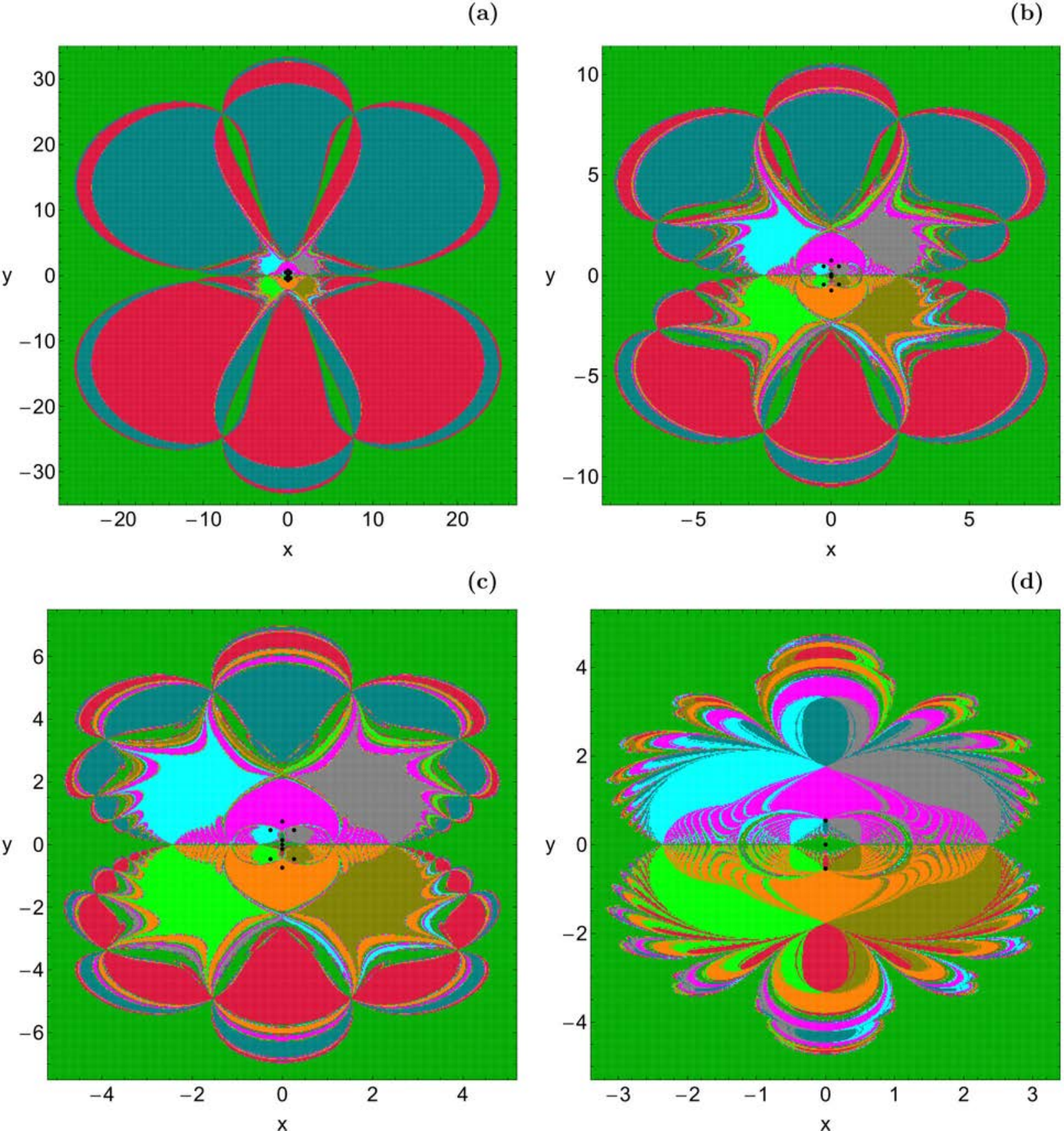}}
\caption{The Newton-Raphson basins of attraction on the configuration $(x,y)$ plane for the fourth case, where nine equilibrium points are present. (a): $\epsilon = 0.5834$; (b): $\epsilon = 0.59$; (c): $\epsilon = 0.62$; (d): $\epsilon = 0.8686$. The positions of the equilibrium points are indicated by black dots. The color code is the same as in Fig. \ref{r1}, while in addition $L_{14}$ (teal) and $L_{15}$ (crimson).}
\label{r4}
\end{figure*}

\begin{figure*}[!t]
\centering
\resizebox{\hsize}{!}{\includegraphics{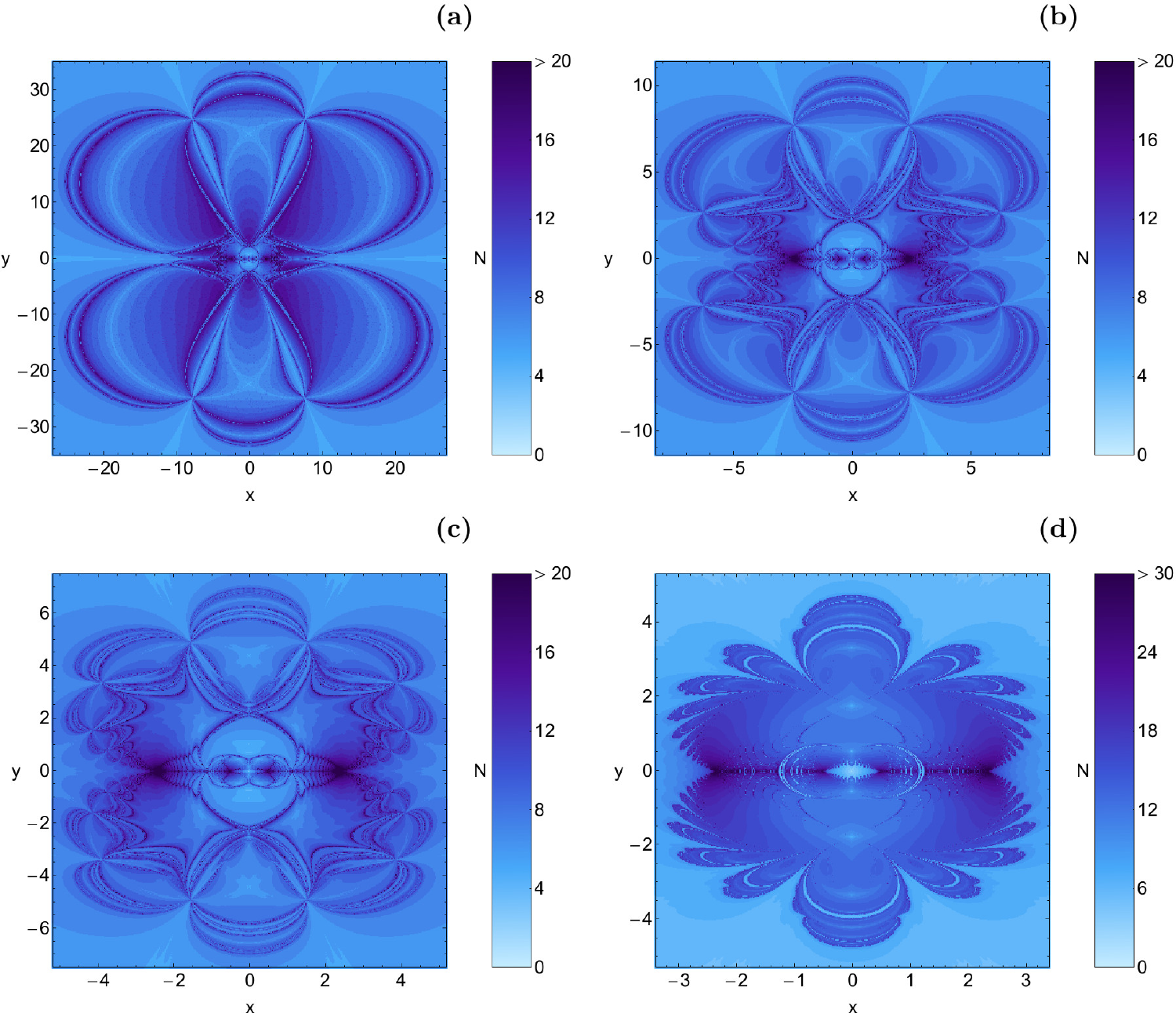}}
\caption{The distribution of the corresponding number $N$ of required iterations for obtaining the Newton-Raphson basins of attraction shown in Fig. \ref{r4}(a-d). The non-converging points are shown in white.}
\label{r4n}
\end{figure*}

\begin{figure*}[!t]
\centering
\resizebox{\hsize}{!}{\includegraphics{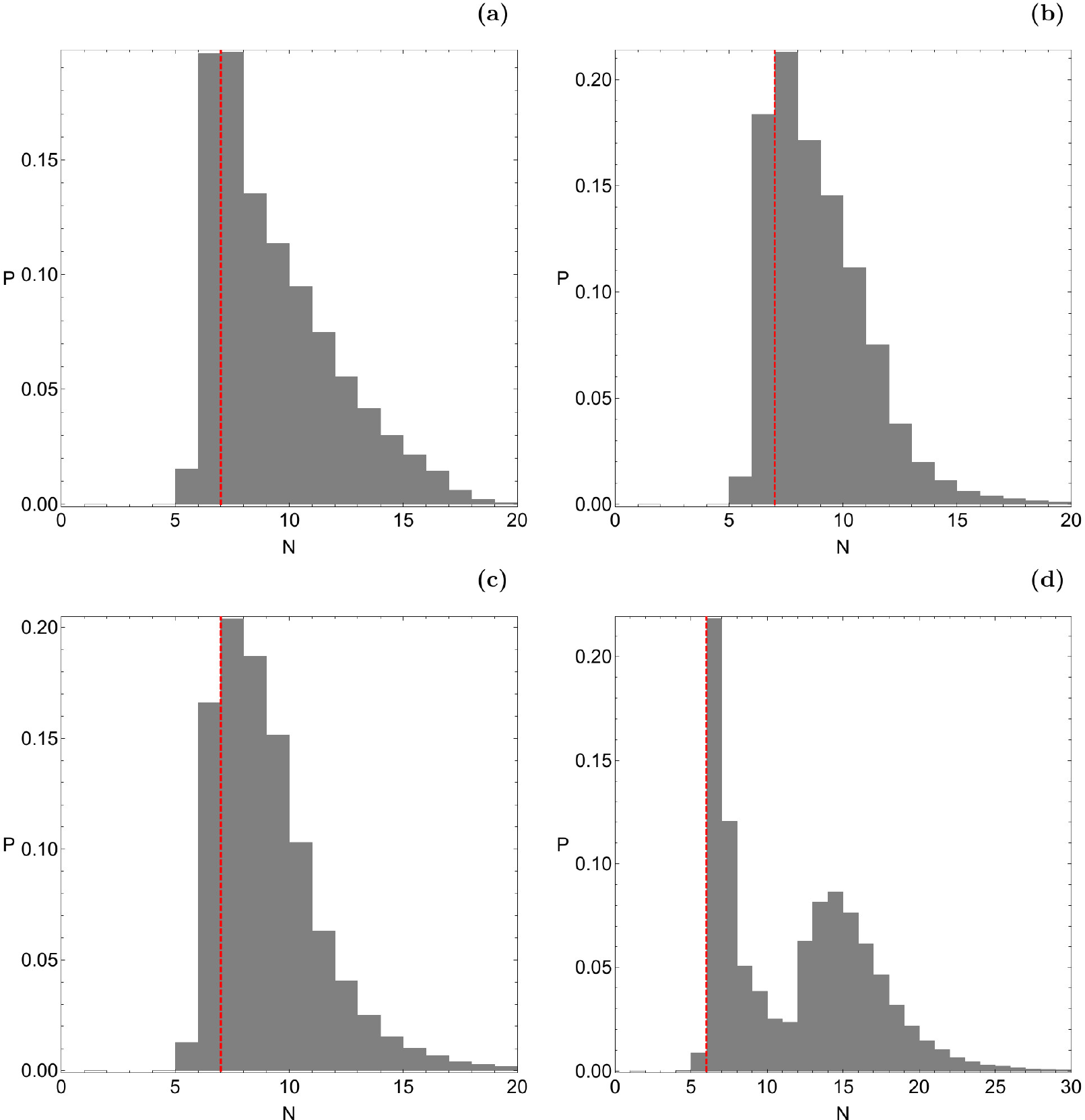}}
\caption{The corresponding probability distribution of required iterations for obtaining the Newton-Raphson basins of attraction shown in Fig. \ref{r4}(a-d). The vertical dashed red line indicates, in each case, the most probable number $N^{*}$ of iterations.}
\label{r4p}
\end{figure*}

\begin{figure*}[!t]
\centering
\resizebox{\hsize}{!}{\includegraphics{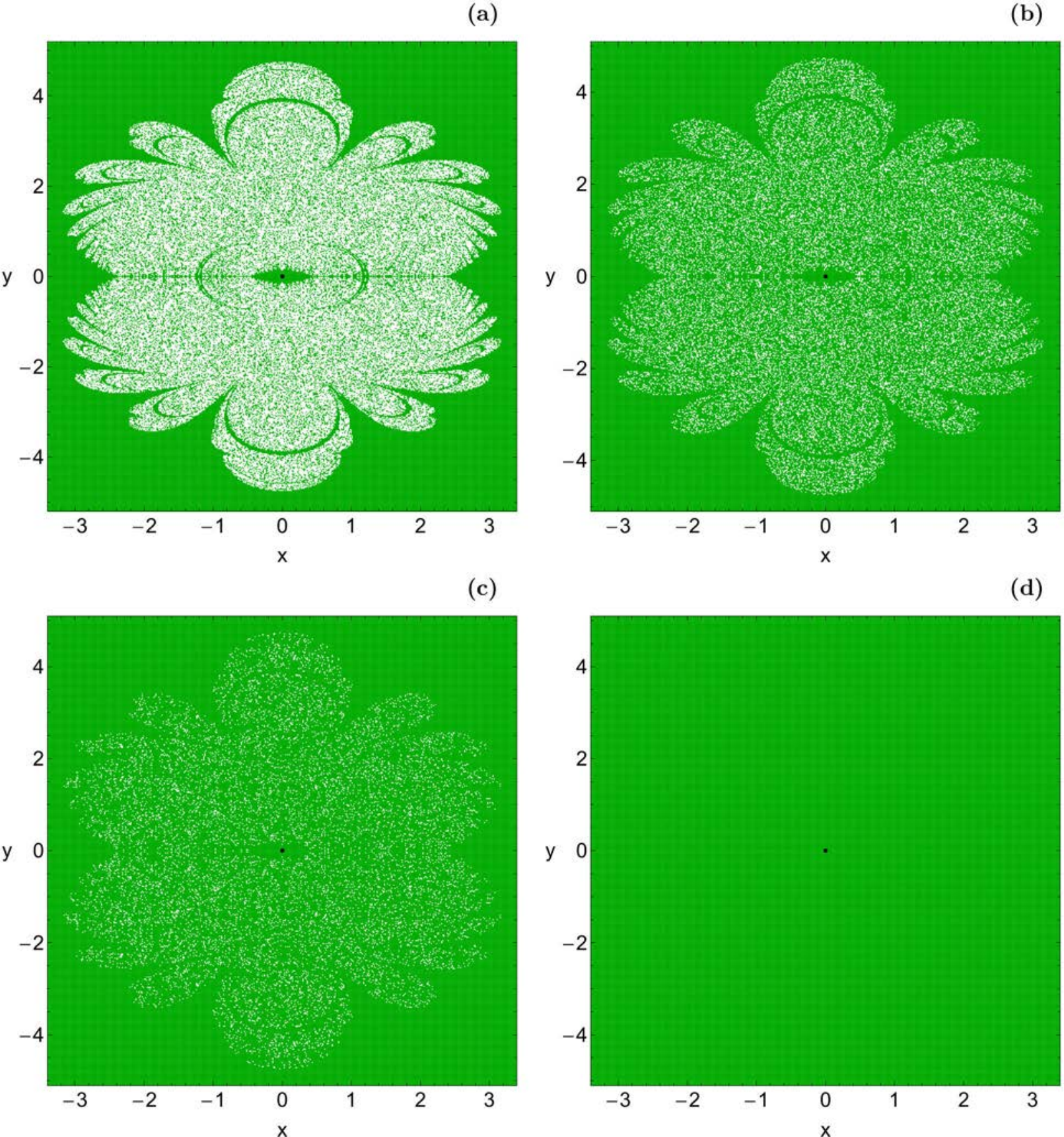}}
\caption{The Newton-Raphson basins of attraction on the configuration $(x,y)$ plane for the fifth case, where only one equilibrium point is present. (a): $\epsilon = 0.8687$; (b): $\epsilon = 0.87$; (c): $\epsilon = 0.872$; (d): $\epsilon = 1$. The positions of the equilibrium points are indicated by black dots. The color code is the same as in Fig. \ref{r1}. White color denotes non-converging points.}
\label{r5}
\end{figure*}

\begin{figure*}[!t]
\centering
\resizebox{\hsize}{!}{\includegraphics{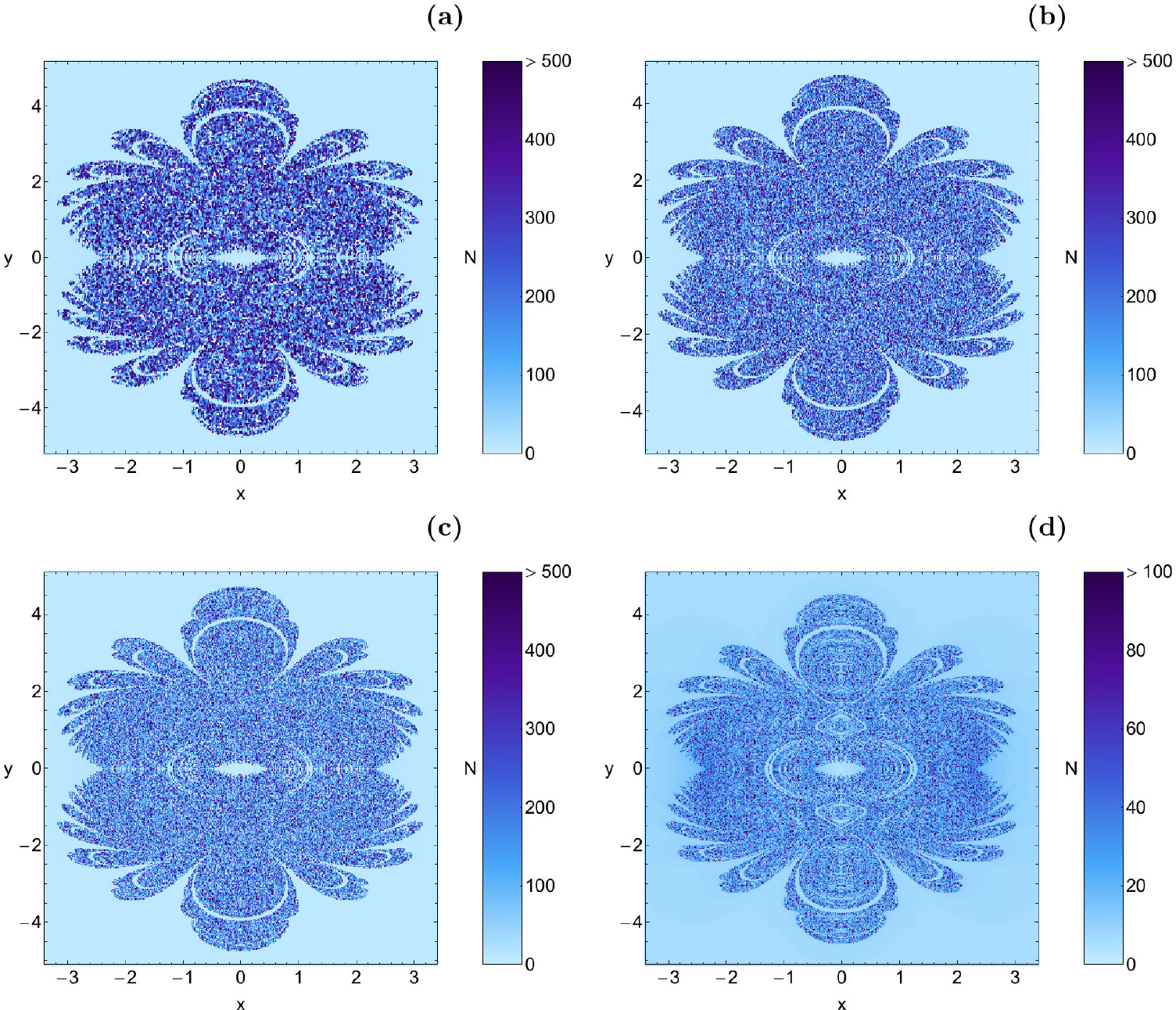}}
\caption{The distribution of the corresponding number $N$ of required iterations for obtaining the Newton-Raphson basins of attraction shown in Fig. \ref{r5}(a-d). The non-converging points are shown in white.}
\label{r5n}
\end{figure*}

\begin{figure*}[!t]
\centering
\resizebox{\hsize}{!}{\includegraphics{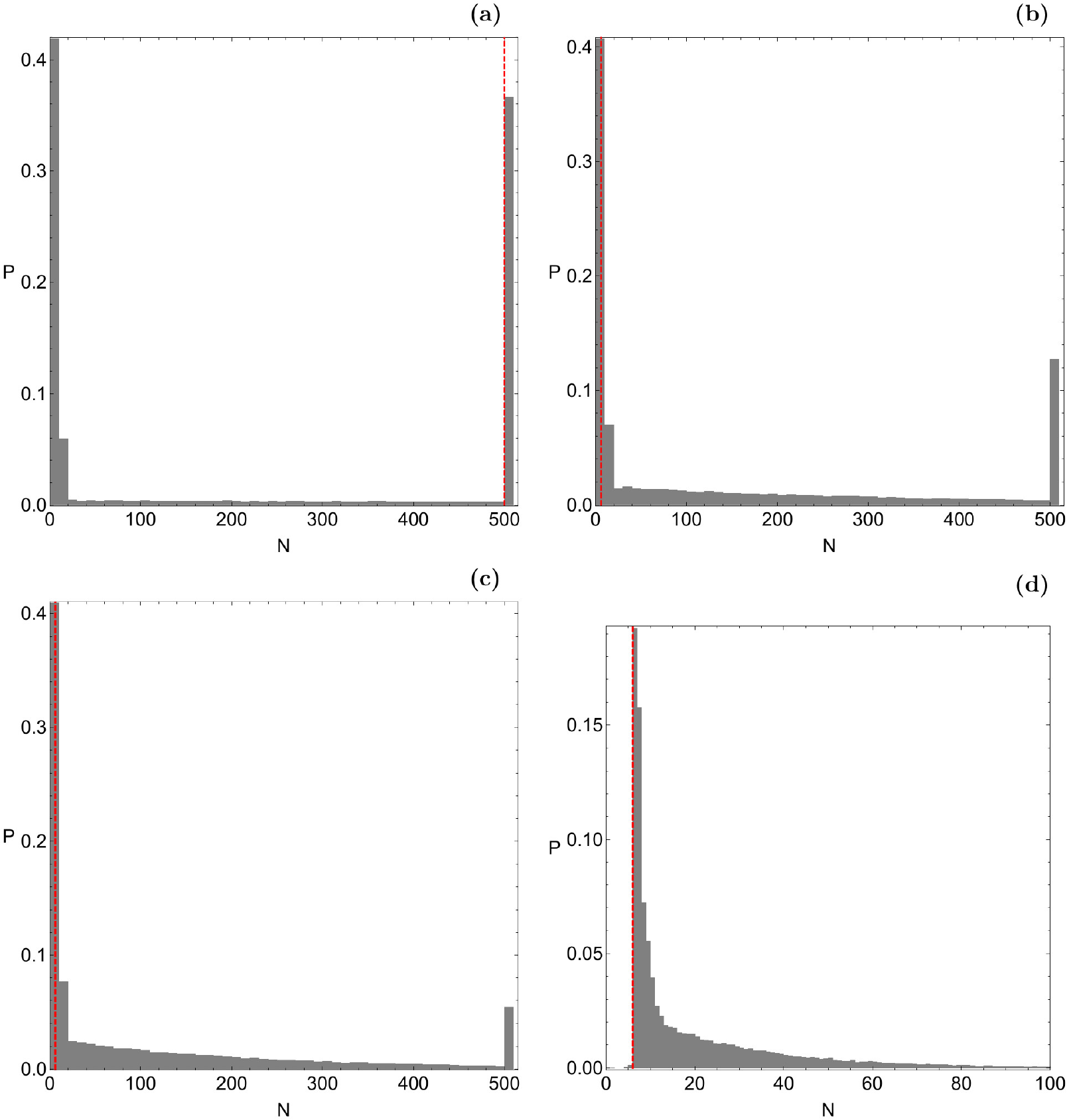}}
\caption{The corresponding probability distribution of required iterations for obtaining the Newton-Raphson basins of attraction shown in Fig. \ref{r5}(a-d). The vertical dashed red line indicates, in each case, the most probable number $N^{*}$ of iterations.}
\label{r5p}
\end{figure*}

\begin{figure*}[!t]
\centering
\resizebox{\hsize}{!}{\includegraphics{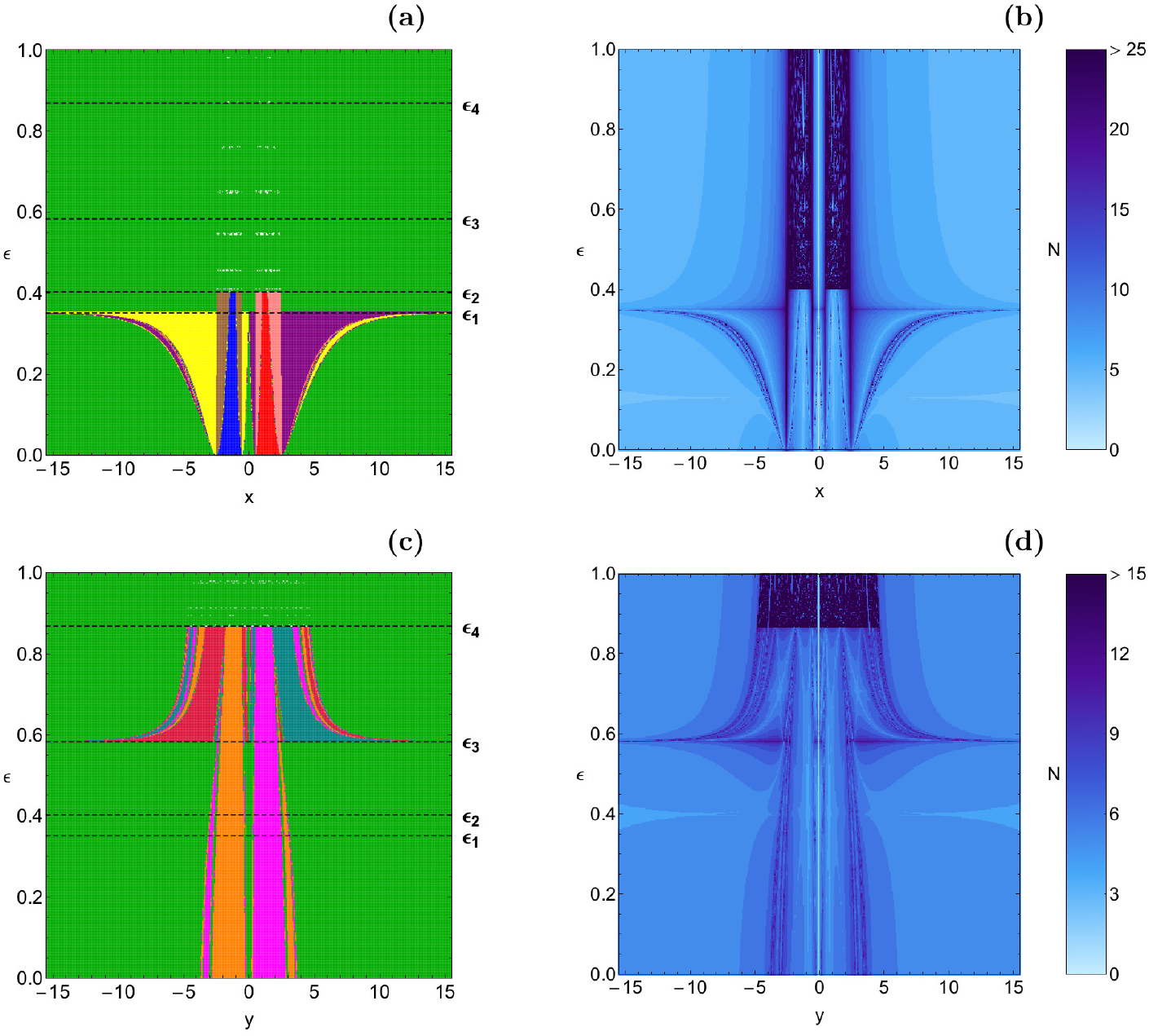}}
\caption{The Newton-Raphson basins of attraction on the (a-upper left): $(x,\epsilon)$ plane and (c-lower left): $(y,\epsilon)$ plane, when $\epsilon \in (0,1]$. The color code denoting the attractors is the same as in Fig. \ref{r4}. The black horizontal dashed lines indicate the four critical values of the transition parameter $\epsilon$. Panels (b) and (d): The corresponding distribution of the required number $N$ of iterations for obtaining the Newton-Raphson basins of convergence shown in panels (a) and (c), respectively.}
\label{xye}
\end{figure*}

The Newton-Raphson basins of convergence when $\epsilon = 0$ (which correspond to the classical Newtonian case) are presented in panel (a) of Fig. \ref{cn}. Different colors are used for each basin of attraction, while the positions of the five equilibrium points (attractors) are indicated by black dots. It is seen that in the case where both primaries have equal masses the axes $x = 0$ and $y = 0$ are axes of symmetry. The distribution of the corresponding number $N$ of iterations is given in panel (b) of the same figure, using tones of blue.

In what follows we will try to determine how the transition parameter $\epsilon$ influences the structure of the Newton-Raphson basins of attraction in the pseudo-Newtonian planar circular restricted three-body problem, by considering five cases regarding the total number of the equilibrium points (attractors). For classifying the initial conditions on the configuration $(x,y)$ plane we will use color-coded diagrams (CCDs), where each pixel is assigned a color, according to the final state (attractor) of the initial condition. Furthermore, the size of the CCDs (or in other words the minimum and the maximum values of the coordinates $x$ and $y$) is controlled in such a way so as to have, in each case, a complete view, regarding the geometry of the structures produced by the attracting domains.

\subsection{Case I: Thirteen equilibrium points}
\label{ss1}

Our numerical exploration begins with the first case where thirteen equilibrium points are present, that is when $0 < \epsilon \leq \epsilon_1$. In Fig. \ref{r1} we present the evolution of the basins of convergence for six values of the transition parameter $\epsilon$. In panel (a), where $\epsilon = 0.01$, it is seen that the structure of the configuration $(x,y)$ plane is almost identical to that observed in Fig. \ref{cn}, for the classical Newtonian case. Nevertheless, one may observe the small attracting domains corresponding to the additional equilibrium points $L_i$, $i = 6,..., 13$. The vast majority of the $(x,y)$ plane is covered by well-formed basins of attraction, while all basin boundaries are highly fractal\footnote{When it is stated that a region is fractal we simply mean that it has a fractal-like geometry, without conducting any additional calculations for computing the fractal dimension as in \citet{AVS01}.}. Thus we may say that these fractal regions behave as chaotic domains. This argument can be justified as follows: for an initial condition $(x_0,y_0)$ inside the chaotic fractal area we will observe that its final state (attractor) is extremely sensitive. More precisely, even a slight deviation in the initial conditions could lead to a completely different final state. Therefore, inside the fractal areas of the configuration $(x,y)$ plane it is next to impossible to predict from which of the attractors (equilibrium points) each initial condition will be attracted by.

The structure of the configuration $(x,y)$ plane changes drastically as the value of the transition parameter increases. In general terms the most noticeable changes are the following:
\begin{itemize}
  \item The extent of the basins of convergence corresponding to libration points $L_2$, $L_3$, $L_4$, and $L_5$ decreases.
  \item The extent of the attracting domains corresponding to equilibrium points $L_6$, $L_9$, $L_{10}$, $L_{11}$, $L_{12}$, $L_{13}$ and especially of $L_7$, $L_8$ increases.
  \item When $\epsilon = 0.3541$ (see panel (f) of Fig. \ref{r1}), that is a value very close to the critical value $\epsilon_1$, the basins of attraction corresponding to the collinear points $L_7$ and $L_8$ dominate, while all other basins, except $L_1$, are confined to the central region of the CCD.
\end{itemize}

Looking carefully at the CCDs presented in Fig. \ref{r1}(a-f) it becomes evident that the basins of attraction corresponding to the central libration point $L_1$ extend to infinity, while on the other hand the extent of all the other basins of attraction is finite. Furthermore, we may say that the shape of the basins of attraction corresponding to equilibrium points $L_2$, and $L_3$ look like exotic bugs with many legs and many antennas, while the shape of the basins of convergence corresponding to all other libration points, except $L_1$, look like butterfly wings.

The distribution of the corresponding number $N$ of iterations is provided, using tones of blue, in Fig. \ref{r1n}(a-f). It is observed that initial conditions inside the attracting regions converge relatively fast $(N < 10)$, while the slowest converging points $(N > 30)$ are those in the vicinity of the basin boundaries. In Fig. \ref{r1p}(a-f) the corresponding probability distribution of iterations is given. The probability $P$ is defined as follows: if $N_0$ initial conditions $(x_0,y_0)$ converge to one of the attractors, after $N$ iterations, then $P = N_0/N_t$, where $N_t$ is the total number of initial conditions in every CCD. It was observed that the most probable number $N^{*}$ of iterations (see the red vertical dashed line in Fig. \ref{r1p}(a-f)) remains almost unperturbed and equal to 6 throughout this region of values of the transition parameter.

\subsection{Case II: Eleven equilibrium points}
\label{ss2}

In this case, where $\epsilon_1 < \epsilon \leq \epsilon_2$, there are eleven equilibrium points: four on the $x$ axis, two on the $y$ axis, four on the $(x,y)$ plane and of course $L_1$ at the center. In Fig. \ref{r2}(a-d) we present the Newton-Raphson basins of convergence for four values of the transition parameter. When $\epsilon = 0.3542$, it is seen in panel (a) of Fig. \ref{r2}, that two sets of thin elongated figure-eight tentacles appear in the vertical direction. With increasing value of $\epsilon$ these tentacles are reduced, while the extent of the basins of attraction corresponding to libration points $L_{10}$, $L_{11}$, $L_{12}$, and $L_{13}$ increases. On the other hand, the extent of all the other attracting domains seems almost unperturbed. We may argue that as value of the transition parameter varies in this interval the geometry of the Newton-Raphson basins of convergence does not change significantly.

The distribution of the corresponding number $N$ of iterations, required for obtaining the desired accuracy in our computations is illustrated in Fig. \ref{r2n}(a-d). Looking at panel (a) of Fig. \ref{r2n} one may observe that the distribution of required iterations, corresponding to the central equilibrium point $L_1$, is very noisy. In other words, for all initial conditions that converge to $L_1$ it is almost impossible to have an estimation about the required number of iterations. This phenomenon becomes much more evident in Fig. \ref{r2p}, where the corresponding probability distribution of iterations is given. Indeed, in panel (a) of Fig. \ref{r2p} we see that the corresponding probability distribution extends up to about $N = 200$, while in all other cases (see panels (b-d) of the same figure) more than 95\% of the initial conditions need less than 35 iterations in order to converge to one of the available attractors. The most probable number $N^{*}$ of iteration is 15 for $\epsilon = 0.3542$, while for all the other studied cases it was found equal to seven.

The strange behavior, regarding the noisy pattern of required iterations, observed for $\epsilon = 0.3542$ can be explained, in a way, as follows: the particular value of the transition parameter is just above the first critical value $\epsilon_1$. Around the critical value $\epsilon_1$ the intrinsic properties of the dynamical system change drastically, as the total number of equilibrium points reduces from thirteen to eleven. We believe that this is exactly the reason of the noisy basin of attraction observed for $\epsilon = 0.3542$.

\subsection{Case III: Seven equilibrium points}
\label{ss3}

Our exploration continues with the third case, where $\epsilon_2 < \epsilon \leq \epsilon_3$. Now there are only seven equilibrium points present. The Newton-Raphson basins of convergence for four values of the transition parameter are depicted in Fig. \ref{r3}(a-d). It is seen that the pattern of panel (a), where $\epsilon = 0.4031$, is almost the same with that observed earlier in panel (d) of Fig. \ref{r2}. The only difference concerns the basins of attraction corresponding to libration points $L_2$ and $L_3$. Now these two points are absent and the corresponding areas on the configuration $(x,y)$ plane are shown in white, which means that these initial conditions do not converge. However additional numerical calculations reveal that these particular initial conditions are in fact slow converging points, which need much more than 500 iteration in order to converge. Moreover it was found that all these slow converging points eventually do converge to the central libration point $L_1$.

As the value of $\epsilon$ increases the pattern of the attracting domain changes. The most important change is the appearance of figure-eight tentacles at the outer parts of the CCDs. These tentacles grow in size (especially along the horizontal direction), while all the other basins of convergence are being confined to the central region of the CCDs (see e.g., panel (d) of Fig. \ref{r3}).

The following Fig. \ref{r3n} shows the distribution of the corresponding number $N$ of iterations. It is interesting to note that the highest numbers of iterations are observed (i) near the vicinity of the places on the $x$ axis, where $L_2$ and $L_3$ used to be and (ii) along the tentacles. The corresponding probability distributions are given in Fig. \ref{r3p}(a-d). It is seen that in all four cases the vast majority of the initial conditions (more than 95\%) converge within the first 35 iterations, while the most probable number of iterations is constant throughout and equal to 7.

\subsection{Case IV: Nine equilibrium points}
\label{ss4}

In the fourth case, where $\epsilon_3 < \epsilon \leq \epsilon_4$, we have the emergence of two new equilibrium points ($L_{14}$ and $L_{15}$) on the vertical $y$ axis. Therefore we have nine libration points in total. The CCDs with the basins of convergence are given in Fig. \ref{r4}(a-d). We observe in panel (a) of Fig. \ref{r4} that extended areas on the configuration $(x,y)$ plane are occupied by the basins of attraction corresponding to $L_{14}$ and $L_{15}$. These basins of convergence have the shape of butterfly wings, while they split into many pieces and their extent is reduced, as we proceed to higher values of the transition parameter. At the same time, the entire pattern of all the attracting domains comes closer to the center. When $\epsilon = 0.8686$ (see panel (d) of Fig. \ref{r4}) the most prominent basins of attraction are those of $L_4$, $L_5$, $L_{10}$, $L_{11}$, $L_{12}$, and $L_{13}$, while those of $L_{14}$ and $L_{15}$ are confined.

In Fig. \ref{r4n} we can see how the numbers $N$ of required iteration are distributed on the configuration $(x,y)$ plane, for the values of $\epsilon$ of Fig. \ref{r4}(a-d). There is no doubt that the most peculiar behavior is observed in panel (d) of Fig. \ref{r4n}, where $\epsilon = 0.8686$, that is a value just below the fourth critical value $\epsilon_4$. More precisely, we observe that all initial conditions composing all basins of attraction, except that of $L_1$, need relatively high numbers of iterations in order to converge, with respect to the required number of iterations for basins of $L_1$. So far we have seen that the highest numbers of iterations correspond mainly to initial conditions in the vicinity of the fractal basin boundaries. However in this case initial conditions of both the fractal basin boundaries and the basins itself require the same high number of iterations. We believe that this strange behaviour must be some kind of intrinsic warning of the dynamical system, thus telling us that something extreme is about to happen. At this point we would like to emphasize that a similar phenomenon (initial conditions inside basins of attraction with large numbers of iterations) has also been observed in the planar equilateral restricted four-body problem (see e.g., panel (i) in Fig. 10 in \cite{Z17a}). In both systems we believe that this behaviour is due to the drastic change of the dynamical properties of the system (change of the total number of equilibrium points). This should be true because in both systems the phenomenon was observed very close to critical values of the parameters, just before the change of the total number of libration points.

Fig. \ref{r4p}(a-d) illustrates the corresponding probability distributions. We see that the most probable number of iterations is 7 for the first three cases, while it drops to 6, when $\epsilon = 0.8686$. In panel (d) of Fig. \ref{r4p} we can see that a second peak appears for $N = 14$. After additional calculations it was revealed that the most probable number $N = 6$ corresponds to initial conditions that converge to the central attractor $L_1$, while the second most probable number $N = 14$ corresponds to initial conditions which converge to all the other attractors.

\subsection{Case V: One equilibrium point}
\label{ss5}

The last case under consideration corresponds to the region $\epsilon_4 < \epsilon \leq 1$, where only the central equilibrium point $L_1$ survives. In Fig. \ref{r5} we present, through the corresponding CCDs, the Newton-Raphson basins of convergence for four values of the transition parameter. In panel (a) of Fig. \ref{r5} one may observe something very interesting as well as very unexpected. About half of the CCD is occupied by initial conditions that do not converge to the attractor. What is simply amazing is the fact that the shape of the non-converging pattern is exactly the same as that shown earlier in panel (d) of Fig. \ref{r4}, where we have measured the highest numbers of iterations. It is as if someone has removed all the points of panel (d) of Fig. \ref{r4} that converge to any attractor other than $L_1$. This behavior is completely new and to our knowledge it has not been observed to any other dynamical system in the past.

The natural question that immediately rises is the following: are these points true non-converging points? In order to answer this question we increased the maximum allowed number of iterations from 500 to 10000 and we reconstructed the CCD. Our results suggest that now all the initial conditions converge, sooner or later, to the central equilibrium point. Therefore, once more we have the case of slow (or even extremely slow) converging points. We believe that that was the extreme change for which the system has informed us earlier, when we have observed high numbers of iterations for initial conditions forming basins of attraction. As the value of the transition parameter $\epsilon$ increases the amount of slow converging points constantly reduces (see panels (b-c) of Fig. \ref{r5}) and when $\epsilon > 0.92$ there is no numerical evidence of slow converging points, whatsoever.

The corresponding distributions of the required number $N$ of iterations and the probabilities $P$ are given in Figs. \ref{r5n}(a-d) and \ref{r5p}(a-d), respectively. Combining the information of these two types of diagrams we can extract two important features: (a) as long as slow-converging points exist the required number of iterations cover all the available interval $N \in [0,500]$, while on the other hand for $\epsilon > 0.92$, where the slow converging points disappear, more than 95\% of the initial conditions converge to $L_1$ within the first 100 iterations, and (b) even when $\epsilon = 1$ the distribution of the required number of iterations $N$ form a specific pattern on the configuration $(x,y)$ plane. This pattern is the almost the same with that of panel (d) of Fig. \ref{r4n}. Thus we may argue that this pattern (which is formed initially when nine attractors are present) is imprinted also in the case where only one attractor exists. As for the most probable number of iterations it remains constant to 6, apart obviously from the first case $(\epsilon = 0.8687)$ where it is equal to 500, due to the large amount of slow converging points.

\subsection{An overview analysis}
\label{over}

Even though the CCDs, on the configuration $(x,y)$ plane, provide sufficient information about the basins of convergence they have a major disadvantage since the information corresponds to a single value of the transition parameter, each time. For eliminating this drawback we have to work on an other type of a two-dimensional plane which will give us the ability to scan a continuous spectrum of values of $\epsilon$. The most convenient way is to set one of the $(x,y)$ coordinates equal to zero and therefore work on the $(x,\epsilon)$ and $(y,\epsilon)$ planes. In panels (a) and (c) of Fig. \ref{xye} we provide the CCDs with the basins of attraction on the $(x,\epsilon)$ and $(y,\epsilon)$ plane, respectively, when $\epsilon \in (0,1]$. Panels (b) and (d) of the same figure contain the corresponding distributions of the required number $N$ of iterations. In both types of planes the four critical values of the transition parameter, $\epsilon_i$, $i = 1,...,4$, are indicated using black horizontal dashed lines.

The CCDs presented in panels (a) and (c) of Fig. \ref{xye} give an excellent perspective regarding the several types of basins of attraction and how they are formed (begin and end) between the critical values of the transition parameter. In both types of planes we detected a small portion (less than 0.1\%) of non-converging initial conditions. Our numerical analysis indicates that the vast majority of these initial conditions are true non-converging points. This must be true because they do not converge, to any of the available attractors, even after $10^6$ iterations. Perhaps, if we increase the maximum allowed number of iterations to an extremely high limit, these initial conditions might converge. Nevertheless, for the time being, we assume that these initial conditions are true non-converging points. Another interesting aspect concerns the required number of iterations. Indeed, for $\epsilon > \epsilon_2$ for the $(x,\epsilon)$ plane, and for $\epsilon > \epsilon_4$, for the $(y,\epsilon)$ plane, near the center there is a considerable amount of initial conditions with relatively high values of iterations. Additional numerical computations (not shown here) suggest that the most probable number of iterations is equal to 5, in both types of planes.

\section{Discussion and conclusions}
\label{conc}

The aim of this work was to numerically compute the basins of attraction, associated with the libration points, in the pseudo-Newtonian planar circular restricted three-body problem, where the primaries have equal masses. Of paramount importance was the determination of the influence of the transition parameter $\epsilon$ on the position as well as on the stability of the equilibrium points. Using the multivariate Newton-Raphson iterative scheme we managed to reveal the beautiful structures of the basins of convergence on several types of two-dimensional planes. The role of the attracting domains is very important since they describe how each initial condition is attracted by the equilibrium points of the system, which act as attractors. Our numerical investigation allowed us to monitor the evolution of the geometry of the basins of convergence as a function of the transition parameter. Moreover, the basins of attraction have been successfully related with both the corresponding distributions of the number of required iterations, and the probability distributions.

As far as we know, there are no previous studies on the Newton-Raphson basins of convergence in the pseudo-Newtonian planar circular restricted three-body problem. Therefore, all the presented numerical outcomes of the current thorough and systematic analysis are novel and this is exactly the importance and the contribution of our work.

The most important outcomes of our numerical analysis can be summarized as follows:
\begin{enumerate}
  \item The transition parameter strongly influences the dynamical properties of the system. Varying its value in the interval $[0,1]$ it was found that the total number of the equilibrium points changes drastically as several points collide with each other and disappear, while in other cases new libration points appear.
  \item The vast majority of the equilibrium points remain either stable or unstable throughout the interval $[0,1]$. Only the libration points $L_1$, $L_4$, and $L_5$ change from stable to unstable, and vice versa, during specific intervals.
  \item It was observed that all types of two-dimensional planes contain a complicated mixture of attracting domains with highly fractal basin boundaries. In the vicinity of the basin boundaries, where the degree of fractality is high, it is almost impossible to know beforehand the final state of an initial condition.
  \item In all examined cases, regarding the numerical value of the transition parameter $\epsilon$, the basins of attraction corresponding to the central equilibrium point $L_1$ extend to infinity. On the other hand, the areas of the basins of convergence associated with all the other libration points are always finite.
  \item In some cases during the scanning of the configuration $(x,y)$ plane we detected a portion of non-converging initial conditions, especially just above the critical value $\epsilon_4$. Additional numerical calculation (by setting a much higher limit of allowed iterations) revealed that these initial conditions are in fact (extremely) slow converging points, corresponding to attractor $L_1$.
  \item Our analysis regarding the convergence properties of the $(x,\epsilon)$ and $(y,\epsilon)$ planes reported the existence of a small amount of non-converging points. In this case, it was found that these particular initial conditions must be true non-converging points since they do not converge, to any of the available attractors, even after $10^6$ numerical iterations.
  \item In the configuration $(x,y)$ plane the most probable number of required iterations, $N^{*}$, was found to mainly vary between 6 and 7 (except of course for values of $\epsilon$ just above the critical levels), while in the $(x,\epsilon)$ and $(y,\epsilon)$ planes it was slightly reduced to 5.
\end{enumerate}

For all the calculation, regarding the determination of the basins of attraction, we used a double precision numerical code, written in standard \verb!FORTRAN 77! \citep{PTVF92}. Furthermore, the latest version 11.1 of Mathematica$^{\circledR}$ \citep{W03} was used for creating all the graphical illustration of the paper. For the classification of each set of initial conditions, in all types of two-dimensional planes, we needed about 6 minutes of CPU time, using a Quad-Core i7 2.4 GHz PC.

We hope that the present numerical outcomes to be useful in the active field of basins of convergence in dynamical systems. Since our present exploration, regarding the attracting domains in the pseudo-Newtonian planar circular restricted three-body problem, was encouraging it is in our future plans to expand our investigation. In particular, it would be of great interest to try other types of iterative formulae (i.e., of higher order, with respect to the classical iterative method of Newton-raphson) and determine how they influence the geometry of the basins of convergence.



\section*{Compliance with Ethical Standards}
\footnotesize

\begin{itemize}
  \item Funding: The author states that he has not received any research grants.
  \item Conflict of interest: The author declares that he has no conflict of interest.
\end{itemize}

\end{document}